\documentclass[onecolumn,draftcls,10pt]{IEEEtran}
\usepackage{amsmath,amssymb,amsthm}\allowdisplaybreaks[1]
\usepackage{cite,bm,pict2e,graphicx}
\usepackage{footmisc}
\newcommand{\E}{\mathsf{E}}
\newcommand{\Var}{\mathsf{Var}}
\newcommand{\Cov}{\mathsf{Cov}}
\newcommand{\tr}{\mathsf{tr}\,}
\newcommand{\diag}{\mathsf{diag}\,}
\newcommand{\eps}{\varepsilon}

\newcommand{\R}{\mathbb{R}}
\newcommand{\A}{\mathbf{A}}
\newcommand{\B}{\mathbf{B}}
\newcommand{\C}{\mathbf{C}}
\newcommand{\Y}{\mathbf{Y}}
\newcommand{\Z}{\mathbf{Z}}
\newcommand{\I}{\mathbf{I}} 
\newcommand{\J}{\mathbf{J}} 
\newcommand{\K}{\mathbf{K}} 
\renewcommand{\k}{\mathbf{k}} 
\newcommand{\0}{\bm{0}} 
\renewcommand{\d}{\partial} 
\newcommand{\dptheta}{\frac{\partial p_\theta}{\partial\theta}}

\newtheorem{theorem}{Theorem}
\newtheorem{corollary}{Corollary}
\newtheorem{lemma}{Lemma}
\newtheorem{proposition}{Proposition}

\renewcommand{\QED}{\QEDopen}
\renewenvironment{proof}[1][Proof]{%
\noindent\hspace{2em}{\itshape #1: }}{%
\hspace*{\fill}~\QED\par\endtrivlist\unskip}

\begin{document}
\title{Information Theoretic Proofs\\ of Entropy Power Inequalities}
\author{Olivier Rioul,~\IEEEmembership{Member, IEEE}\thanks{A summary of these results is to be presented at the IEEE Symposium on Information Theory 2007 in Nice, France.}\\
Institut T\'el\'ecom\\
T\'el\'ecom ParisTech\\
CNRS LTCI\\
Paris, France\\
\texttt{olivier.rioul@telecom-paristech.fr}
\date{August 23, 2010}}
\maketitle%\IEEEpeerreviewmaketitle

\begin{abstract}
While most useful information theoretic inequalities can be deduced from the basic properties of entropy or mutual information, up to now Shannon's entropy power inequality (EPI) is an exception: Existing information theoretic proofs of the EPI hinge on representations of differential entropy using either Fisher information or minimum mean-square error (MMSE), which are derived from de Bruijn's identity. In this paper, we first present an unified view of these proofs, showing that they share two essential ingredients: 1) a data processing argument applied to a covariance-preserving linear transformation; 2) an integration over a path of a continuous Gaussian perturbation. Using these ingredients, we develop a new and brief proof of the EPI through a mutual information inequality, which replaces Stam and Blachman's Fisher information inequality (FII) and an inequality for MMSE by Guo, Shamai and Verd\'u used in earlier proofs. The result has the advantage of being very simple in that it relies only on the basic properties of mutual information.
These ideas are then generalized to various extended versions of the EPI: Zamir and Feder's generalized EPI for linear transformations of the random variables, Takano and Johnson's EPI for dependent variables, Liu and Viswanath's covariance-constrained EPI, and Costa's concavity inequality for the entropy power.
\end{abstract}

\begin{keywords}
Entropy power inequality (EPI), differential entropy, mutual information, data processing inequality, Fisher information inequality (FII), Fisher information, de Bruijn's identity, minimum mean-square error (MMSE), relative entropy, divergence.
\end{keywords}

\section{Introduction}

In his 1948 historical paper, Shannon proposed the \emph{entropy power inequality} (EPI)\cite[Thm. 15]{Shannon48}, which asserts that the entropy power of the sum of independent random vectors is at least the sum of their entropy powers; equality holds iff\footnote{if and only if.} the random vectors are Gaussian with proportional covariances. 
The EPI is one of the deepest inequalities in information theory, and has a long history.
Shannon gave a variational argument\cite[App. 6]{Shannon48} to show that 
the entropy of the sum of two independent random vectors of given entropies has a stationary point where the two random vectors are Gaussian with proportional covariance matrices, but this does not exclude the possibility that the stationary point is not a global minimum. 
Stam\cite{Stam59} credits de Bruijn with a first rigorous proof of the EPI in the case where at most one of the random vectors is \emph{not} Gaussian, using a relationship between differential entropy and Fisher information now known as \emph{de Bruijn's identity}.
A general proof of the EPI is given by Stam\cite{Stam59} (see also Blachman\cite{Blachman65}), based on a related \emph{Fisher information inequality} (FII). Stam's proof is simplified in\cite{DemboCoverThomas91} and\cite{CarlenSoffer91}.
Meanwhile, Lieb\cite{Lieb78} proved the EPI via a strengthened Young's inequality from functional analysis. While Lieb's proof does not use information theoretic arguments, Dembo, Cover and Thomas\cite{DemboCoverThomas91} showed that it can be recast in a unified proof of the EPI and the Brunn-Minkowski inequality in geometry (see also\cite{CostaCover84, GuleryuzLutwakYang02}), which was included in the textbook by Cover and Thomas\cite[\S~17.8]{CoverThomas06}. 
Recently, Guo, Shamai and Verd\'u\cite{GuoShamaiVerdu05} found an integral representation of differential entropy using minimum mean-square error (MMSE), which yields another proof of the EPI\cite{VerduGuo06, GuoShamaiVerdu06a}. A similar, continuous-time proof via causal MMSE was also proposed by Binia~\cite{Binia07}.
The original information theoretic proofs (by Stam and Blachman, and by Verd\'u, Guo and Shamai) were first given for scalar random variables, and then generalized to the vector case either by induction on the dimension~\cite{Stam59,Blachman65} or by extending the required tools~\cite{DemboCoverThomas91,VerduGuo06}.

The EPI is used to bound capacity or rate-distortion regions for certain types of channel or source coding schemes, especially to prove converses of coding theorems in the case where optimality cannot be resolved by Fano's inequality alone. Shannon used the EPI as early as his 1948 paper\cite{Shannon48} to bound the capacity of non-Gaussian additive noise channels. Other examples include Bergmans' solution\cite{Bergmans74} to the scalar Gaussian broadcast channel problem, generalized to the multiple-input multiple-output (MIMO) case in\cite{WeingartenSteinbergShamai06,MohseniCioffi06}; Leung-Yan Cheong and Hellman's determination of the secrecy capacity of the Gaussian wire-tap channel\cite{Leung-Yan-CheongHellman78}, extended to the multiple access case in\cite{TekinYener06,TekinYener07}; Costa's solution to the scalar Gaussian interference channel problem\cite{Costa85a}; Ozarow's solution to the scalar Gaussian source two-description problem\cite{Ozarow80}, extended to multiple descriptions at high resolution in\cite{Zamir99}; and Oohama's determination of the rate-distortion regions for various multiterminal Gaussian source coding schemes\cite{Oohama97,Oohama98,Oohama05,Oohama06}.
% including separate coding of two correlated Gaussian memoryless sources\cite{Oohama97}, the quadratic Chief Executive Officer (CEO) problem\cite{Oohama98}, and the many-help-one problem\cite{Oohama05,Oohama06}.
It is interesting to note that in all the above applications, the EPI is used only in the case where all but one of the random vectors in the sum are Gaussian.
The EPI for general independent random variables, as well as the corresponding FII, also find application in blind source separation and deconvolution in the context of independent component analysis (see, e.g.,\cite{Donoho81, BercherVignat00,VrinsVerleysen05}), and is instrumental in proving a strong version of the central limit theorem with convergence in relative entropy\cite{Barron86,Brown82,CarlenSoffer91,Johnson00,JohnsonSuhov01,ABBN04a,JohnsonBarron04}.

It appears that the EPI is perhaps the only useful information theoretic inequality that is \emph{not} proved through basic properties of entropy or mutual information.
In this paper, we fill the gap by providing a new proof, with the following nice features: 
\begin{itemize}
\item it hinges solely on the elementary properties of Shannon's mutual information,
sidestepping both Fisher's information and MMSE. Thus, it relies only on the most basic principles of information theory;
\item it does not require scalar or vector identities such as de Bruijn's identity, nor integral representations of differential entropy;
\item the vector case is handled just as easily as the scalar case, along the same lines of reasoning; and
\item it goes with a mutual information inequality (MII), which has its own interest.
\end{itemize}
Before turning to this proof, we make a detailed analysis of the existing information theoretic proofs\footnote{Lieb's excepted, since it belongs to mathematical analysis and cannot be qualified as an ``information theoretic'' proof.}  of the EPI.
The reasons for this presentation are as follows:
\begin{itemize}
\item it gives some idea of the level of difficulty that is required to understand conventional proofs. The new proof presented in this paper is comparatively simpler and shorter;
\item it focuses on the essential ingredients common to all information theoretic proofs of the EPI, namely data processing inequalities and integration over a path of continuous Gaussian perturbation. This serves as a insightful guide to understand the new proof which uses the same ingredients, though in an more expedient fashion;
\item it simplifies some of the conventional argumentation and provides intuitive interpretations for the Fisher information and de Bruijn's identity, which have their own interests and applications. In particular, a new, simple proof of a (generalized) de Bruijn's identity, based on a well-known estimation theoretic relationship between relative entropy and Fisher information, is provided;
\item it offers a unified view of the apparently unrelated existing proofs of the EPI. They do not only share essentials, but can also be seen as variants of the same proof;
and
\item it derives the theoretical tools that are necessary to further discuss the relationship between the various approaches, especially for extended versions of the EPI.
\end{itemize}
The EPI has been generalized in various ways. Costa\cite{Costa85} (see also \cite{Dembo89}) strengthened the EPI for two random vectors in the case where one of these vectors is Gaussian, by showing that the entropy power is a concave function of the power of the added Gaussian noise. Zamir and Feder\cite{ZamirFeder93, ZamirFeder93a, ZamirFeder93b} generalized the scalar EPI by considering the entropy power of an arbitrary linear transformation of the random variables. Takano\cite{Takano96} and Johnson\cite{Johnson04} provided conditions under which the original EPI still holds for two dependent variables. Recently, Liu and Viswanath\cite{LiuViswanath06,LiuViswanath07} generalized the EPI by considering a covariance-constrained optimization problem motived by multiterminal coding problems.
The ideas in the new proof of the EPI presented in this paper are readily extended to all these situations. Again, in contrast to existing proofs, the obtained proofs rely only on the basic properties of entropy and mutual information. In some cases, further generalizations of the EPI are provided.

The remainder of this paper is organized as follows.
We begin with some notations and preliminaries. Section~\ref{sec-earlier} surveys earlier information theoretic proofs of the EPI and presents a unified view of the proofs.
Section~\ref{sec-epi} gives the new proof of the EPI, along with some discussions and perspectives. The reader may wish to skip directly to the proof in this section, which does not use the tools presented earlier.
Section~\ref{sec-zfepi} extends the new proof to Zamir and Feder's generalized EPI for arbitrary linear transformations of independent variables.
Section~\ref{sec-tjepi} adapts the new proof to the case of dependent random vectors, generalizing the results of Takano and Johnson.
Section~\ref{sec-lvepi} generalizes the new proof to an explicit formulation of Liu and Viswanath's EPI under a covariance constraint, based on the corresponding MII.
Section~\ref{sec-cepi} gives a proof of the concavity of the entropy power (Costa's EPI) based on the MII, which relies only on the properties of mutual information.
Section~\ref{sec-open} concludes this paper with some open questions about a recent generalization of the EPI to arbitrary subsets of independent variables\cite{ABBN04,MadimanBarron06,MadimanBarron07,TulinoVerdu06} and a collection of convexity inequalities for linear ``gas mixtures''.

\subsection{Notations} 
In this paper, to avoid $\log e$ factors in the derivations, information quantities are measured in \emph{nats}---we shall use only natural logarithms and exponentials.
Random variables or vectors are denoted by upper case letters, and their values denoted by lower case letters. The expectation $\E(\cdot)$ is taken over the joint distribution of the random variables within the parentheses. 
The covariance matrix of a random (column) $n$-vector $X$ is 
$\Cov(X)=\E\bigl((X-\E(X))(X-\E(X))^t\bigr)$,
and its variance is the trace of the covariance matrix: 
$\Var(X)=\tr\bigl(\Cov(X)\bigr)=\E\bigl(\|X-\E(X)\|^2\bigr)$.
We also use the notation $\sigma^2_X=\frac{1}{n}\Var(X)$
for the variance per component. We say that $X$ is \emph{white} if its covariance matrix is proportional to the identity matrix, and \emph{standard} if it has unit covariance matrix $\Cov(X)=\I$.

With the exception of the conditional mean $\E(X|Y)$, which is a function of $Y$, all quantities in the form $f(X|Y)$ used in this paper imply expectation over $Y$, following the usual convention for conditional information quantities.
Thus the conditional covariance matrix is $\Cov(X|Y)=\E\bigl((X-\E(X|Y))(X-\E(X|Y))^t\bigr)$, and the conditional variance is $\Var(X|Y) = \tr \bigl( \Cov(X|Y) \bigr)= \E \bigl( \|X-\E(X|Y) \|^2\bigr)$, that is,  the MMSE in estimating $X$ given the observation $Y$, achieved by the conditional mean estimator $\hat{X}(Y)=\E(X|Y)$.

The diagonal matrix with entries $a_i$ is denoted by $\diag(a_i)_i$.
We shall use the partial ordering between real symmetric matrices where $\A\leq\B$ means that the difference is positive semidefinite, that is, for any real vector $x$, $x^t\A x\leq x^t\B x$. 
Clearly $\A\leq\B$ implies $\C\A\C\leq \C\B\C$ for any symmetric matrix $\C$, and $\B^{-1}\leq \A^{-1}$ if $\A$ and $\B$ are invertible and $\A$ is positive semidefinite.

Given a function $f(x)$, $\frac{\d f}{\d x}$ denotes the gradient, a (column) vector of partial derivatives $(\frac{\d f}{\d x_i})_i$, and $\frac{\d^2 f}{\d x^2}$ denotes the Hessian, a matrix of second partial derivatives $(\frac{\d^2f}{\d x_i\d x_j})_{i,j}$. 
We shall use Landau's notations $o(f)$ (a function which is negligible compared to $f$ in the neighborhood of some limit value of $x$) and $O(f)$  (a function which is dominated by $f$ in that neighborhood).

\subsection{Definition of the differential entropy}\label{sec-hdef}

Let $X$ be any random $n$-vector having probability density $p(x)$ (with respect to the Lebesgue measure). Its (differential) entropy is defined by 
\begin{equation}\label{de}
h(X) = \E \log \frac{1}{p(X)} = -\int p(x)\log p(x) \,dx
\end{equation}
provided that this integral exists in the generalized sense---that is, the positive and negative parts of this integral are not both infinite. Thus we may have $h(X)=+\infty$ if $\E \log^{+} \frac{1}{p(X)}=+\infty$ and $\E \log^{-} \frac{1}{p(X)}<+\infty$; and $h(X)=-\infty$ if $\E \log^{+} \frac{1}{p(X)}<+\infty$ and $\E \log^{-} \frac{1}{p(X)}=+\infty$, where we have noted $x^{+}=\max(x,0)$ and $x^{-}=\max(-x,0)$.

The differential entropy is not always well defined. Take, for example, $p(x)=\frac{1}{2x\log^{2}x}$ for $0<x<1/e$ and $e<x<+\infty$, and $p(x)=0$ otherwise. In this case it is easy to check that both positive and negative parts of the integral $\int p(x)\log p(x) \,dx$ are infinite.
In spite of that differential entropy is frequently encountered in the literature, the author was unable to find simple, general conditions under which it is well defined. An exception is reference~\cite{Vajda89} which gives the sufficient condition that $p^{\alpha}(x)$ is Lebesgue-integrable for any $\alpha$ in the range $\alpha_{0}\leq \alpha\leq 2$ where $0<\alpha_{0}<1$. The following result may be more useful for practical considerations.

\begin{proposition}[Well Defined Entropy]\label{prop-hdef}
If $\E \bigl(\log(1+\|X\|)\bigr)$ is finite, in particular if $X$ has finite first or second moments, then $h(X)$ is well defined and is such that $-\infty \leq h(X) <+\infty$.
\end{proposition}

\begin{proof}
It is sufficient to prove that the positive part $h^{+}(X)=\E \log^{+} \frac{1}{p(X)}$ of~\eqref{de} is finite.
Let $q(x)$ be the Cauchy density defined by
\begin{equation}
q(x)=\frac{\Gamma(\frac{n+1}{2})}{\pi^{\frac{n+1}{2}}} \frac{1}{(1+\|x\|^{2})^{\frac{n+1}{2}}}.
\end{equation}
Since $u\log u\geq -1/e$ for all $u>0$, we have
\begin{subequations}
\begin{align}
h^{+}(X) &= -\int_{0<p(x)\leq 1} p(x)\log p(x) dx \\
&=-\int_{0<p(x)\leq 1} p(x)\log q(x) dx + \int_{0<p(x)\leq 1} q(x) \frac{p(x)}{q(x)}\log \frac{q(x)}{p(x)} dx\\
&\leq \log \frac{\pi^{\frac{n+1}{2}}}{\Gamma(\frac{n+1}{2})} + \frac{n+1}{2} \E \bigl(\log(1+\|X\|^{2})\bigr) + \frac{1}{e} \int_{p(x)\leq 1} q(x) dx\\
&\leq \log \frac{\pi^{\frac{n+1}{2}}}{\Gamma(\frac{n+1}{2})} +(n+1) \E \bigl(\log(1+\|X\|)\bigr) + \frac{1}{e}
\end{align}
\end{subequations}
which is finite by assumption.
\end{proof}
It is easy to adapt the proof in the particular case where $\E(\|X\|)$ or $\E(\|X\|^{2})$ is finite by letting $q(x)$ be an exponential Laplacian or normal Gaussian distribution, respectively. The proof can also be shorten slightly by applying the theorem of Gel'fand-Yaglom-Perez~\cite[chap.~2]{Pinsker64} to the relative entropy $D(p\|q)=\int p(x) \log \frac{p(x)}{q(x)} \,dx$, which is finite because $X$ is absolutely continuous with respect to the measure defined by density $q(x)$.

In the situation of Proposition~\ref{prop-hdef} it is sometimes convenient to extend the definition by setting $h(X)=-\infty$ when $X$ does \emph{not} admit a density with respect to the Lebesgue 
measure---in particular, when the distribution of $X$ has a probability mass assigned to one or more singletons in $\R^n$ (see, e.g.~\cite{Costa85} and~\cite[p.~6]{JohnsonBook04}). This convention can be justified by a limiting argument in several cases. Another justification appears in Lemma~\ref{lem-entropy} below.

\subsection{Entropy-Power Inequalities (EPI)}

The \emph{entropy power} $N(X)$ of a random $n$-vector $X$ with (differential) entropy $h(X)$ is\cite{Shannon48}
\begin{equation}\label{ep}
N(X) = \frac{e^{\frac{2}{n}h(X)}}{2\pi e}
\end{equation}
In the following, we assume that entropies are well defined, possibly with value  $h(X)=-\infty$ or $N(X)=0$. 
The scaling properties
\begin{equation}\label{episcale}
\begin{aligned}
h(aX)&=h(X)+n\log |a|\\
N(aX)&=a^2N(X),
\end{aligned}
\end{equation}
where $a\in\R$, follow from the definitions by a change of variable argument.

Suppose $X$ has finite covariances. The \emph{non-Gaussianness} of $X$ is the relative entropy (divergence) with respect to a Gaussian random vector $X^*$ with identical second moments:
\begin{equation}\label{nongaussH}
D(X\|X^*)=h(X^*)-h(X)
\end{equation}
where $h(X^*)=\frac{1}{2}\log \bigl( (2\pi e)^n |\Cov(X)|\bigr)$.
With the convention of the preceding section, one has $D(X\|X^*)=+\infty$ if $h(X)=-\infty$.
Since~\eqref{nongaussH} is nonnegative and vanishes iff $X$ is Gaussian, the entropy power~\eqref{ep} satisfies the inequalities
\begin{equation}\label{shannonineq}
N(X) \leq |\Cov(X)|^{1/n} \leq \sigma^2_X,
\end{equation}
with equality in the first inequality iff $X$ is Gaussian, and in the second iff $X$ is white.
In particular, $N(X)$ is the power of a white Gaussian random vector having the same entropy as $X$. 

From these observations, it is easily found that Shannon's EPI can be given several equivalent forms:
\begin{proposition}[Equivalent EPIs]\label{prop-epi}
The following inequalities, each stated for finitely many independent random vectors $(X_i)_i$ with finite differential entropies, and real-valued coefficients $(a_i)_i$, are equivalent.
\begin{subequations}\label{epi}
\begin{align}\label{epi1}
N(\sum_i a_i X_i) &\geq \sum_i a_i^2 N(X_i),
\\\label{epi2}
h(\sum_i a_i X_i) &\geq h(\sum_i a_i \widetilde{X}_i),
\\\label{epi3}
h(\sum_i a_i X_i) &\geq \sum_i a_i^2 h(X_i) \qquad (\sum_i a_i^2 = 1),
\end{align}
\end{subequations}
where the $(\widetilde{X}_i)_i$ are independent Gaussian random vectors with proportional covariances (e.g., white) and corresponding entropies $h(\widetilde{X}_i)=h(X_i)$.
\end{proposition}
We have presented weighted forms of the inequalities to stress the similarity between~\eqref{epi1}--\eqref{epi3}. Note that by \eqref{episcale}, the normalization $\sum_i a_i^2 = 1$ is unnecessary for \eqref{epi1} and~\eqref{epi2}. 
The proof is given in\cite{DemboCoverThomas91} and is also partly included in\cite{VerduGuo06} in the scalar case. For completeness we include a short proof\footnote{This proof corrects a small error in\cite{DemboCoverThomas91}, namely, that the first statement in the proof of Theorem~7 in\cite{DemboCoverThomas91} is false when the Gaussian random vectors do not have identical covariances.}.
  
\begin{proof}
That \eqref{epi1}, \eqref{epi2} are equivalent follows from the equalities $\sum_i a_i^2 N(X_i) = \sum_i a_i^2 N(\widetilde{X_i}) = N(\sum_i a_i \widetilde{X_i})$. To prove that~\eqref{epi3} is equivalent to \eqref{epi1} we may assume that $\sum_i a_i^2 = 1$. Taking logarithms of both sides of~\eqref{epi1}, inequality~\eqref{epi3} follows from the concavity of the logarithm. Conversely, taking exponentials of both sides of~\eqref{epi3}, inequality~\eqref{epi1} follows provided that the $(X_i)_i$ have equal entropies. But the latter condition is unnecessary because if~\eqref{epi1} is satisfied for the random vectors $(N(X_i)^{-1/2}X_i)_i$ of equal entropies, then upon modification of the coefficients it is also satisfied for the $(X_i)_i$.
\end{proof}

Inequality~\eqref{epi1} is equivalent to the classical formulation of the EPI\cite{Shannon48} by virtue of the scaling property~\eqref{episcale}.
Inequality~\eqref{epi2} is implicit in\cite[App.~6]{Shannon48}, where Shannon's line of thought is to show that the entropy of the sum of independent random vectors of given entropies has a minimum where the random vectors are Gaussian with proportional covariance matrices. It was made explicit by~Costa and Cover\cite{CostaCover84}. Inequality~\eqref{epi3} is due to Lieb\cite{Lieb78} and is especially interesting since all available proofs of the EPI are in fact proofs of this inequality. It can be interpreted as a concavity property of entropy\cite{DemboCoverThomas91} under the covariance-preserving transformation
\begin{equation}\label{transformation}
(X_i)_i \longmapsto Y=\sum_i a_i X_i \qquad (\sum_i a_i^2 = 1).
\end{equation} 

Interestingly,~\eqref{epi3} is most relevant in several applications of the EPI. Although the preferred form for use in coding applications\cite{Bergmans74,WeingartenSteinbergShamai06, MohseniCioffi06,Leung-Yan-CheongHellman78,TekinYener06,TekinYener07,Costa85a ,Ozarow80, Zamir99,Oohama97,Oohama98,Oohama05,Oohama06} is the inequality $N(X+Z)\geq N(X)+N(Z)$, where $Z$ is Gaussian independent of~$X$, 
Liu and Viswanath\cite{LiuViswanath06,LiuViswanath07} suggest that the EPI's main contribution to multiterminal coding problems is for solving optimization problems of the form $\max_{X} h(X)-\mu h(X+Z)$,
%where the maximization is over all random vectors $X$ independent of $Z$. 
whose solution is easily determined from the convexity inequality~\eqref{epi3} as shown in Section~\ref{sec-lvepi}.
Also,~\eqref{epi3} is especially important for solving blind source separation and deconvolution problems, because it implies that negentropy $c=-h$ satisfies the requirements for a ``contrast function'':
\begin{equation}\label{contrast}
c(\sum_i a_i X_i)\leq \max_i c(X_i)\qquad (\sum_i a_i^2 = 1),
\end{equation}
which serves as an objective function to be maximized in such problems\cite{Donoho81,BercherVignat00,VrinsVerleysen05}. Finally, the importance of the EPI for proving strong versions of the central limit theorem is through~\eqref{epi3} interpreted as a monotonicity property of entropy for standardized sums of independent variables\cite{Barron86,ABBN04a}.

\section{Earlier Proofs Revisited}  \label{sec-earlier}
  
 \subsection{Fisher Information Inequalities (FII)}

Conventional information theoretic proofs of the EPI use an alternative quantity, the Fisher information (or a disguised version of it), for which the statements corresponding to \eqref{epi} are easier to prove.
The \emph{Fisher information matrix} $\J(X)$ of a random $n$-vector $X$ with density $p(x)$ is\cite{DemboCoverThomas91,CoverThomas06}
\begin{equation}\label{fim}
\J(X) = \Cov\bigl(S(X)\bigr)
\end{equation}
where the zero-mean random variable (log-derivative of the density)
\begin{equation}\label{score}
S(X) = \nabla \log p(X) = \frac{\nabla p(X)}{p(X)}
\end{equation}
is known as the \emph{score}.
The \emph{Fisher information} $J(X)$ is the trace of~\eqref{fim}:
\begin{equation}\label{fi}
J(X) = \Var\bigl(S(X)\bigr) = \E \frac{\|\nabla p(X)\|^2}{p(X)^2}.
\end{equation}
In this and the following subsections, we assume that probability densities are sufficiently smooth with sufficient decay at infinity so that Fisher informations exist,
%\footnote{See e.g., \cite{Stam59} for a more precise formulation.}, 
possibly with the value  $J(X)=+\infty$. % if $X$ has a probability mass assigned to one or more singletons in $\R^n$. 
The scaling properties
\begin{equation}\label{fiscale}
\begin{aligned}
S(aX)&=a^{-1}S(X)\\
J(aX)&=a^{-2}J(X)
\end{aligned}
\end{equation}
follow from the definitions by a change of variable argument.
Note that if $X$ has independent entries, then $\J(X)$ is the diagonal matrix $\J(X)=\diag \bigl(J(X_i)\bigr)_i$. 

It is easily seen that the score $S(X)$ is a linear function of $X$ iff $X$ is Gaussian.
Therefore, a measure of \emph{non-Gaussianness} of $X$ is the mean-square error of the score with respect to the (linear) score $S^*$ of a Gaussian random vector $X^*$ with identical second moments:
\begin{equation}\label{nongaussJ}
\E \bigl(\|S(X)-S^*(X)\|^2\bigr)=J(X)-J(X^*)
\end{equation}
where $J(X^*)= \tr \bigl(\Cov(X)^{-1}\bigr)$.
Since \eqref{nongaussJ} is nonnegative and vanishes iff $X$ is Gaussian, the Fisher information~\eqref{fi} satisfies the inequalities
\begin{equation}\label{cramerrao}
J(X) \geq \tr \bigl(\Cov(X)^{-1}\bigr) \geq \frac{n}{\sigma^2_X}.
\end{equation}
The first inequality (an instance of the Cram\'er-Rao inequality) holds with equality iff $X$ is Gaussian, while the second inequality (a particular case of the Cauchy-Schwarz inequality on the eigenvalues of $\Cov(X)$) holds with equality iff $X$ is white.
In particular, $nJ^{-1}(X)$ is the power of a white Gaussian random vector having the same Fisher information as $X$. 

%Stam's FII can be given three equivalent forms.
\begin{proposition}[Equivalent FIIs]\label{prop-fii}
The following inequalities, each stated for finitely many independent random vectors $(X_i)_i$ with finite Fisher informations, and real-valued coefficients $(a_i)_i$, are equivalent.
\begin{subequations}\label{fii}
\begin{align}\label{fii1}
J^{-1}(\sum_i a_i X_i) &\geq \sum_i a_i^2 J^{-1}(X_i),
\\\label{fii2}
J(\sum_i a_i X_i) &\leq J(\sum_i a_i \widetilde{X}_i),
\\\label{fii3}
J(\sum_i a_i X_i) &\leq \sum_i a_i^2 J(X_i) \qquad (\sum_i a_i^2 = 1),
\end{align}
\end{subequations}
where the $(\widetilde{X}_i)_i$ are independent Gaussian random vectors with proportional covariances (e.g., white) and corresponding Fisher informations $J(\widetilde{X}_i)=J(X_i)$.
\end{proposition}
There is a striking similarity with Proposition~\ref{prop-epi}. The proof is the same, with the appropriate changes---the convexity of the hyperbolic $1/x$ is used in place of the concavity of the logarithm---and is omitted. Inequality~\eqref{fii3} is due by Stam and its equivalence with~\eqref{fii1} was pointed out to him by de Bruijn\cite{Stam59}.
It can be shown\cite{Papathanasiou93,Zamir98} that the above inequalities also hold for positive semidefinite symmetric matrices,
where Fisher informations~\eqref{fi} are replaced by Fisher information matrices~\eqref{fim}.

Similarly as for~\eqref{epi3}, inequality~\eqref{fii3} can be interpreted as a convexity property of Fisher information\cite{DemboCoverThomas91} under the covariance-preserving transformation~\eqref{transformation}, or as a monotonicity property for standardized sums of independent variables\cite{Barron86,JohnsonBarron04}. 
It implies that the Fisher information $C=J$ satisfies~\eqref{contrast}, and therefore, can be used as a contrast function in deconvolution problems~\cite{Donoho81}.
The FII has also been used to prove a strong version of the central limit theorem\cite{Barron86,Brown82,CarlenSoffer91,Johnson00,JohnsonSuhov01,ABBN04a,JohnsonBarron04} and a characterization of the Gaussian distribution by rotation\cite{Itoh89,Papathanasiou93}.

%, and in parametric estimation from filtered measurements\cite{Zamir98}. % and Zamir 1995, proc. (it is better...)

\subsection{Data Processing Inequalities for Least Squares 
Estimation\protect\footnote{We use the term ``least squares estimation'' for any estimation procedure based on the mean-squared error criterion.}}\label{sec-dpi}

Before turning to the proof the FII, it is convenient and useful to make some preliminaries about data processing inequalities for Fisher information and MMSE. In estimation theory, the importance of the Fisher information follows from the Cram\'er-Rao bound (CRB)\cite{CoverThomas06} on the mean-squared error of an estimator of a parameter $\theta\in\R^m$ from a measurement $X\in\R^n$. In this context, $X$ is a random $n$-vector whose density $p_\theta(x)$ depends on $\theta$, and the (parametric) \emph{Fisher information matrix} is defined by\cite{CoverThomas06,Zamir98}
\begin{equation}\label{pfim}
\J_\theta (X) =\Cov\bigl(S_\theta(X)\bigl)
\end{equation}
where $S_\theta(X)$ is the (parametric) \emph{score function},
\begin{equation}\label{pscore}
S_\theta(X)= \frac{\partial}{\partial\theta} \log p_\theta(X).
\end{equation}
In some references the parametric Fisher information is defined as the trace of~\eqref{pfim}:
\begin{equation}\label{pfi}
J_\theta(X) = \Var\bigl(S_\theta(X)\bigr).
\end{equation}
In the special case where $\theta\in\R^n$ is a \emph{translation parameter}: $p_\theta(x)=p(x+\theta)$, we recover the earlier definitions~\eqref{fim}--\eqref{fi}: $S(X)=S_\theta(X-\theta)$, $\J(X)=\J_\theta(X-\theta)$, and $J(X)=J_\theta(X-\theta)$. More generally, it is easily checked that for any $a\in\R$, 
\begin{subequations}
\begin{align}
S_\theta(X-a\theta)&=aS(X)\label{scorea}\\
J_\theta(X-a\theta)&=a^2J(X) \label{fia}.
\end{align}
\end{subequations}

The optimal unbiased estimator of $\theta$ given the observation $X$, if it exists, is such that the mean-square error meets the CRB (reciprocal of the Fisher information)\cite{CoverThomas06}. Such an optimal estimator is easily seen to be a linear function of the score~\eqref{pscore}. Thus it may be said that the score function $S_\theta(X)$ represents the optimal least squares estimator of $\theta$.
When the estimated quantity $\theta$ is a random variable (i.e., not a parameter), the optimal estimator is the conditional mean estimator $\E(\theta|X)$ and the corresponding miminum mean-square error (MMSE) is the conditional variance $\Var(\theta|X)$.

In both cases, there is a \emph{data processing theorem}\cite{Zamir98} relative to a transformation $X\to Y$ in a \emph{Markov chain} $\theta\to X\to Y$, that is, for which $Y$ given $X$ is independent of $\theta$. The emphasize the similarity between these data processing theorems and the corresponding quantities of Fisher information and MMSE, we first prove the following ``chain rule'', which states that the optimal estimation given $Y$ of $\theta$ results from the optimal estimation given $Y$ of the optimal estimation given $X$ of $\theta$:

\begin{proposition}[Data Processing Theorem for Estimators]\label{prop-dpt}
If $\theta\to X\to Y$ form a Markov chain, then
\begin{subequations}
\begin{align}
\E(\theta|Y) &= \E \bigl(\E(\theta|X)|Y\bigr)\label{dpt}\\
S_\theta(Y) &= \E_\theta \bigl(S_\theta(X)|Y\bigr).\label{pdpt}
\end{align}
\end{subequations}
\end{proposition}

\begin{proof}
In the nonparametric case the Markov chain condition can written as $p(\theta|x,y)=p(\theta|x)$. Multiplying by $\theta\, p(x|y)$ gives $\theta \, p(\theta,x|y)=\theta\, p(\theta|x)p(x|y)$, which integrating over $\theta$ and $x$ yields~\eqref{dpt}.
In the parametric case the Markov chain condition can be written as $p_\theta(x,y)=p_\theta(x)p(y|x)$ where the distribution $p(y|x)$ is independent of $\theta$. Differentiating with respect to $\theta$ gives $\frac{\d p_\theta}{\d\theta}(x,y)=\frac{\d p_\theta}{\d\theta}(x)p(y|x)$; dividing by $p_\theta(y)$ and applying Bayes' rule yields the relation
$\dptheta(x,y)/p_\theta(y) = \dptheta(x)/p_\theta(x) \, p_\theta(x|y)$, which integrating over $x$ yields~\eqref{pdpt}.
\end{proof}

From Proposition~\ref{prop-dpt} we obtain a unified proof of the corresponding \emph{data processing inequalities} for least squares estimation, which assert that the transformation $X\to Y$ reduces information about $\theta$, or in other words, that no clever transformation can improve the inferences made on the data measurements: compared to $X$, the observation $Y$ yields a worse estimation of $\theta$.
\begin{proposition}[Estimation Theoretic Data Processing Inequalities]\label{prop-dpi}
If $\theta\to X\to Y$ form a Markov chain, then
\begin{subequations}
\begin{align}
\Cov(\theta|Y) &\geq \Cov(\theta|X) \label{dpimmsem}\\
\J_\theta(Y) &\leq \J_\theta(X).\label{dpifim}
\end{align}
\end{subequations}
In particular,
\begin{subequations}
\begin{align}
\Var(\theta|Y) &\geq \Var(\theta|X) \label{dpimmse}\\
J_\theta(Y) &\leq J_\theta(X).\label{dpifi}
\end{align}
\end{subequations}
Equality holds iff
\begin{subequations}
\begin{align}
\E(\theta|X)&=\E(\theta|Y) \text{ a.e.,}\label{dpimmseeq}\\
S_\theta(X)&=S_\theta(Y) \text{ a.e.,}\label{dpifieq}
\end{align}
\end{subequations}
respectively.
\end{proposition}
\begin{proof}
The following identity (``total law of covariance'') is well known and easy to check:
\begin{equation}\label{tlc}
\Cov(U)=\Cov(U|V)+\Cov\bigl(\E(U|V)\bigr).
\end{equation}
For $U=\E(\theta|X)$ or $U=S_\theta(X)$, and $V=Y$, we obtain, by Proposition~\ref{prop-dpt},
\begin{subequations}
\begin{align}
\Cov(\theta|X)&=\Cov(\theta|Y) - \Cov\bigl(\E(\theta|X)|Y\bigr)\label{dpimmsemloss}\\
\J_\theta(X)&=\J_\theta(Y) + \Cov\bigl(S_\theta(X)|Y\bigr).\label{dpifimloss}
\end{align}
\end{subequations}
where in deriving~\eqref{dpimmsemloss} we have also used~\eqref{tlc} for $U=\theta$. 
Since covariance matrices are positive semidefinite, this proves~\eqref{dpimmsem},~\eqref{dpifim}, and~\eqref{dpimmse},~\eqref{dpifi} follow by taking the trace.
Equality holds in~\eqref{dpimmsem},~\eqref{dpimmse} or in~\eqref{dpifim},~\eqref{dpifi} iff $E(\theta|X)$ or $S_\theta(X)$ is a deterministic function of $Y$, which by Proposition~\ref{prop-dpt} is equivalent to~\eqref{dpimmseeq} or~\eqref{dpifieq}, respectively.
\end{proof}

Stam\cite{Stam59} mentioned that~\eqref{dpifi} is included in the original work of Fisher, in the case where $Y$ is a deterministic function of $X$. A different proof of~\eqref{dpifim} is provided by Zamir\cite{Zamir98}. The above proof also gives, via~\eqref{dpimmsemloss},~\eqref{dpifimloss} or the corresponding identities for the variance, explicit expressions for the information ``loss'' due to processing. The equality conditions correspond to the case where the optimal estimators given $X$ or $Y$ are the same.
In particular, it is easily checked that~\eqref{dpifieq} is equivalent to the fact that $\theta\to Y \to X$ (in this order) also form a Markov chain, that is, $Y$ is a ``sufficient statistic'' relative to $X$\cite{CoverThomas06}.

As a consequence of Proposition~\ref{prop-dpi} we obtain a simple proof of the following relation between Fisher information and MMSE in the case where estimation is made in Gaussian noise:
\begin{proposition}[Complementary Relation between Fisher Information and MMSE]\label{prop-fimmse}
If $Z$ is Gaussian independent of $X$, then
\begin{equation}\label{fimmse}
\J(X+Z)\Cov(Z)+\Cov(Z)^{-1}\Cov(X|X+Z) = \mathbf{I}
\end{equation}
In particular, if $Z$ is white Gaussian,
\begin{equation}\label{fimmsewhite}
\sigma^2_Z J(X+Z)+\sigma^{-2}_Z \Var(X|X+Z) = n
\end{equation}
\end{proposition}

\begin{proof}
Apply~\eqref{dpifimloss} to the Markov chain $\theta\to (X,Z-\theta) \to X+Z-\theta$, where $X$ and $Z$ are independent of $\theta$ and of each other. Since $S_\theta(X,Z-\theta)=S_\theta(X)+S_\theta(Z-\theta)=S(Z)=-\Cov(Z)^{-1}(Z-\E(Z))$, we have $J_\theta(X,Z-\theta)=J(Z)=\Cov(Z)^{-1}$. Therefore,~\eqref{dpifimloss} reads
$$
\Cov(Z)^{-1} = \J(X+Z) + \Cov(Z)^{-1} \Cov(Z|X+Z) \Cov(Z)^{-1}
$$
Noting that $Z-\E(Z|X+Z)=\E(X|X+Z)-X$, one has $\Cov(Z|X+Z)=\Cov(X|X+Z)$ and~\eqref{fimmse} follows upon multiplication by $\Cov(Z)$. For white Gaussian $Z$,~\eqref{fimmsewhite} follows by taking the trace.
\end{proof}

As noted by Madiman and Barron~Ê\cite{MadimanBarron07}, \eqref{fimmsewhite} is known in Bayesian estimation (average risk optimality): see \cite[Thm. 4.3.5]{LehmannCasella98} in the general situation where $X+Z$ is replaced any variable $Y$ such that $p(y|x)$ belongs to an exponential family parameterized by $x$.
It was rediscovered independently by Budianu and Tong\cite{BudianuTong05}, and by Guo, Shamai and Verd\'u\cite{GuoShamaiVerdu04,GuoShamaiVerdu05}. Relation~\eqref{fimmse} was also rederived by Palomar and Verd\'u~\cite{PalomarVerdu06} as a consequence of a generalized de Bruijn's identity (Corollary~\ref{cor-debruijn} below). 
Other existing proofs are by direct calculation. The above proof is simpler and offers an intuitive alternative based on the data processing theorem.

To illustrate~\eqref{fimmse}, consider the case where $X$ and $Z$ are zero-mean Gaussian. In this case, the conditional mean estimator $\E(X|X+Z)$ is linear of the form $\A(X+Z)$, where $\A$ is given by the Wiener-Hopf equations $\A\Cov(X+Z)=\E(X(X+Z)^t)=\Cov(X)$. Therefore $\E(X|X+Z)=\Cov(X)\Cov(X+Z)^{-1}(X+Z)=X+Z-\Cov(Z)\Cov(X+Z)^{-1}(X+Z)$. This gives, after some calculations, $\Cov(X|X+Z)=\Cov(Z)-\Cov(Z)\Cov(X+Z)^{-1}\Cov(Z)$. But this expression is also an immediate consequence of~\eqref{fimmse} since one has simply $\J(X+Z)=\Cov(X+Z)^{-1}$.

For standard Gaussian $Z$,~\eqref{fimmsewhite} reduces to the identity $J(X+Z)+\Var(X|X+Z) = n$, which constitutes a simple complementary relation between Fisher information and MMSE. The estimation of $X$ from the noisy version $X+Z$ is all the more better as the MMSE is lower, that is, as $X+Z$ has higher Fisher information. Thus Fisher information can be interpreted a measure of least squares (nonparametric) estimation's efficiency, when estimation is made in additive Gaussian noise.

\subsection{Proofs of the FII via Data Processing Inequalities}

Three distinct proofs of the FII~\eqref{fii3} are available in the literature. In this section, we show that these are in fact variations on the same theme: thanks to the presentation of Section~\ref{sec-dpi}, each proof can be easily interpreted as an application of the data processing theorem to the (linear) deterministic transformation $(X_i)_i\mapsto Y$ given by~\eqref{transformation}, or in parametric form:
\begin{equation}\label{ptransformation}
 Y-\theta=\sum_i a_i (X_i-a_i\theta) \qquad (\sum_i a_i^2 = 1).
\end{equation} 
 
\subsubsection{Proof %of the FII~\eqref{fii3} 
via the Data Processing Inequality for Fisher Information}\label{sec-stam}
This is essentially Stam's proof\cite{Stam59} (see also Zamir\cite{Zamir98} for a direct proof of~\eqref{fii1} by this method). Simply apply~\eqref{dpifi} to the transformation~\eqref{ptransformation}:
\begin{equation}\label{pfii}
J_\theta(\sum_i a_i X_i -\theta) \leq J_\theta\bigl((X_i-a_i\theta)_i\bigr) = \sum_i J_\theta(X_i-a_i\theta)
\end{equation}
From~\eqref{fia}, the FII~\eqref{fii3} follows.

\subsubsection{Proof %of the FII~\eqref{fii3} 
via Conditional Mean Representations of the Score} \label{sec-blachman}
This proof is due to Blachman\cite{Blachman65} in the scalar case ($n=1$). His original derivation is rather technical, since it involves a direct calculation of the convolution of the densities of independent random variables $U$ and $V$ to establish that $S(U+V)=\E(\lambda S(U)+(1-\lambda)S(V)|U+V)$ for any $0\leq\lambda\leq 1$, followed by an application of the Cauchy-Schwarz inequality.
The following derivation is simpler and relies on the data processing theorem: By Proposition~\ref{prop-dpt} applied to the transformation~\eqref{ptransformation}, 
\begin{equation*}
S_\theta(\sum_i a_i X_i -\theta) =
 \E\Bigl(S_\theta\bigl((X_i-a_i\theta)_i\bigr) | \sum_i a_i X_i-\theta\Bigr) 
 = \E\bigl(\sum_i S_\theta(X_i-a_i\theta) | \sum_i a_i X_i\Bigr) 
\end{equation*}
which from~\eqref{scorea} gives the following conditional mean representation of the score:
\begin{equation}\label{cmrscore}
S(\sum_i a_i X_i)=\E\bigl(\sum_i a_i S(X_i)|\sum_i a_i X_i\bigr)
\end{equation}
This representation includes Blachman's as a special case (for two variables $U=a_1X_1$ and $V=a_2X_2$). The rest of Blachman's argument parallels the above proof of the data processing inequality for Fisher information (Proposition~\ref{prop-dpi}): His application of the Cauchy-Schwarz inequality\cite{Blachman65} is simply a consequence of the law of total variance $\Var(U)=\Var(U|V)+\Var\bigl(\E(U|V)\bigr)$. Indeed, taking $U=\sum_i a_i S(X_i)$, $V=\sum_i a_i X_i$, and using~\eqref{cmrscore}, the inequality $\Var(U)\geq \Var\bigl(\E(U|V)\bigr)$ reduces to the FII~\eqref{fii3}.
Thus we see that, despite appearances, the above two proofs of Stam and Blachman are completely equivalent.

\subsubsection{Proof %of the FII~\eqref{fii3} 
via the Data Processing Inequality for MMSE} \label{sec-verduguo}
This proof is due to Verd\'u and Guo\cite{VerduGuo06}, which use MMSE in lieu of Fisher's information. Apply~\eqref{dpimmse} to the transformation~\eqref{transformation}, in which each $X_i$ is replaced by $X_i+Z_i$, where the $(Z_i)_i$ are i.i.d. white Gaussian of variance $\sigma^2$. Noting $Z=\sum_i a_i Z_i$, this gives
\begin{equation}\label{mmsei}
\Var(\sum_i a_i X_i | \sum_i a_i X_i +Z)
\geq \Var(\sum_i a_i X_i | (X_i +Z_i)_i)=
\sum_i a^2_i  \Var( X_i | X_i +Z_i)
\end{equation}
where $Z$ is also white Gaussian of variance $\sigma^2$. By the complementary relation~\eqref{fimmsewhite} (Proposition~\ref{prop-fimmse}), this inequality is equivalent to the FII $J(\sum_i a_i X_i + Z) \leq \sum_i a_i^2 J(X_i+Z_i)$ and letting $\sigma^2\to  0$ gives~\eqref{fii3}\footnote{This continuity argument is justified in~\cite{PinskerPrelovMeulen98}.}. Again this proof is equivalent to the preceding ones, by virtue of the complementary relation between Fisher information and MMSE.

\subsubsection{Conditions for Equality in the FII}\label{sec-eqfii}
 The case of equality in~\eqref{fii3} was settled by Stam\cite{Stam59} and Blachman\cite{Blachman65}. In Stam's approach, by Proposition~\ref{prop-dpi}, equation~\eqref{dpifieq}, equality holds in~\eqref{pfii} iff $\sum_i S_\theta(X_i-a_i\theta) = S_\theta(\sum_i a_i X_i -\theta)$,
that is, using~\eqref{scorea},
\begin{equation}\label{scoreidentity}
\sum_i a_i S(X_i) = S(\sum_i a_i X_i) \text{ a.e.}
\end{equation}
This equality condition is likewise readily obtained in Blachman's approach above.
Obviously, it is satisfied only if all scores for which $a_i\ne 0$ are \emph{linear} functions, which means that equality holds in the FII only if the corresponding random vectors are \emph{Gaussian}. 
In addition, replacing the scores by their expressions for Gaussian random $n$-vectors in~\eqref{scoreidentity}, it follows easily by identification that these random vectors have identical covariance matrices. Thus equality holds in~\eqref{fii3} iff all random vectors $X_i$ such that $a_i\ne 0$ are Gaussian with identical covariances.

Verd\'u and Guo do not derive the case of equality in\cite{VerduGuo06}. From the preceding remarks, however, it follows that equality holds in~\eqref{mmsei} only if the $(X_i+Z_i)_i$ for which $a_i\ne 0$ are Gaussian---and therefore, the corresponding $(X_i)_i$ are themselves Gaussian. This result is not evident from estimation-theoretic properties alone in view of the equality condition~\eqref{dpimmseeq} in the data processing inequality for the MMSE.

\subsection{De Bruijn's Identity}\label{sec-debruijn}

\subsubsection{Background}
De Bruijn's identity is the fundamental relation between differential entropy and Fisher information, and as such, is used to prove the EPI~\eqref{epi3} from the corresponding FII~\eqref{fii3}. This identity can be stated in the form\cite{DemboCoverThomas91}
\begin{equation}\label{debruijnstandard}
\frac{d}{dt}h(X+\sqrt{t}\, Z)\Bigr|_{t=0}=\frac{1}{2}J(X)
\end{equation}
where $Z$ is standard Gaussian, independent of the random $n$-vector $X$. 
It is proved in the scalar case in\cite{Stam59}, generalized to the vector case by Costa and Cover\cite{CostaCover84} and to nonstandard Gaussian $Z$ by Johnson and Suhov\cite{JohnsonSuhov01,Johnson04}.
The conventional, technical proof of de Bruijn's identity relies on a diffusion equation satisfied by the Gaussian distribution and is obtained by integrating by parts in the scalar case and invoking Green's identity in the vector case. We shall give a simpler and more intuitive proof of a generalized identity for arbitrary (not necessarily Gaussian) $Z$:

\begin{proposition}[De Bruijn's Identity]\label{prop-debruijn}
For any two independent random $n$-vectors $X$ and $Z$ such that $\J(X)$ exists and $Z$ has finite covariances,
\begin{subequations}\label{debruijn}
\begin{equation}\label{debruijngeneral}
\frac{d}{dt} h(X+\sqrt{t}\, Z) \Bigr|_{t=0} = \frac{1}{2}\,\tr\bigl(\J(X)\,\Cov(Z)\bigr).
\end{equation}
In particular, if $Z$ is white or $X$ has i.i.d. entries, 
\begin{equation}\label{debruijnwhite}
\frac{d}{dt} h(X+\sqrt{t}\, Z) \Bigr|_{t=0} = \frac{1}{2}\sigma^2_Z \, J(X).
\end{equation}
\end{subequations}
\end{proposition}

\subsubsection{A Simple Proof of De Bruijn's Identity}

The proof is based on the following observation.
Setting $\theta=\sqrt{t}$,~\eqref{debruijngeneral} can be rewritten as a first-order Taylor expansion in $\theta^2$:
\begin{equation}\label{debruijnexpansion}
h(X+\theta Z)-h(X) = \frac{\theta^2}{2}\,\E\Bigl( \bigl(Z-\E(Z)\bigr)^t \J(X)\bigl(Z-\E(Z)\bigr)\Bigr) + o(\theta^2).
\end{equation}
Now, there is a well-known, similar expansion of  \emph{relative entropy} (divergence) 
\begin{equation}\label{divergence}
D_X(p_\theta\|p_{\theta'}) =  \E_\theta \log \frac{p_\theta(X)}{p_{\theta'}(X)} 
\end{equation}
in terms of \emph{parametric} Fisher information~\eqref{pfim}, for a parameterized family of densities $p_\theta(x)$, $\theta\in\mathbb{R}^m$. Indeed, since the divergence is nonnegative and vanishes for $\theta'=\theta$, its second-order Taylor expansion takes the form~\cite{Kullback68}
\begin{equation}\label{kullback}
D_X(p_\theta\|p_{\theta'}) =\frac{1}{2}\, (\theta'-\theta)^t \J_\theta(X)  (\theta'-\theta) + o(\|\theta'-\theta\|^2),
\end{equation}
where $\J_\theta (X)$ is the positive semidefinite Hessian matrix of the divergence, that is, $\J_\theta (X)=\frac{\d^2}{\d{\theta'}^2}D_X(p_\theta\|p_{\theta'})\Bigr|_{\theta'=\theta}\!\!=\E_\theta \frac{\d^2}{\d{\theta}^2} \log \frac{1}{p_\theta(X)}$, which is easily seen to coincide with definition~\eqref{pfim}\footnote{Even though the divergence is not symmetric in $(\theta,\theta')$, it is locally symmetric in the sense that~\eqref{kullback} is also the second-order Taylor expansion for $D_X(p_{\theta'}\|p_{\theta})$.}.
In view of the similarity between~\eqref{debruijnexpansion} and~\eqref{kullback}, the following proof of de Bruijn's identity is almost immediate.

\begin{proof}[Proof of Proposition~\ref{prop-debruijn}]
Let $Y=X+\theta Z$ and write mutual information $I(X+\theta\, Z;Z)=h(X+\theta\, Z)-h(X)$ as a conditional divergence: $I(Y,Z)=D\bigl(p(y|z)\|p(y)\bigr)=\E \bigl(D(p_X(y-\theta Z)\| p_Y(y))$. Making the change of variable $u=y-\theta z$ gives
$
I(X+\theta Z; Z) =\E_Z\bigl( D(q_0\|q_\theta) \bigr)
$,
where $q_\theta(u)=p_{X+\theta Z}(u+\theta z)$ is the parameterized family of densities  of a random variable $U$, and $q_0(u)=p_X(u)$. 
Therefore, by~\eqref{kullback} for scalar $\theta$,
\begin{equation}\label{kullback2}
I(X+\theta Z; Z) = \frac{\theta^2}{2} \E_Z\bigl( J_0(U) \bigr) + o(\theta^2),
\end{equation}
where $J_0(U)$ is the parametric Fisher information of $U$ about $\theta=0$, which is easily determined as follows.

Expanding $p(y|z)=p_X(y-\theta z)$ about $\theta=0$ gives $p(y|z)=p_X(y)-\theta z^t\nabla p_X(y)+o(\theta)$, and therefore, $p(y)=\E (p(y|Z))=p_X(y)-\theta\, \E(Z)^t\nabla p_X(y)+o(\theta)$, where the limit for $\theta\to 0$ and the expectation have been exchanged,  due to Lebesgue's convergence theorem and the fact that  $Z$ has finite covariances. It follows that $q_\theta(u)=p_{Y}(u+\theta z)=q_0(u)+\theta\, \bigl(z-\E(Z)\bigr)^t \nabla p_X(u)+o(\theta)$ so that the (parametric) score of $U$ for $\theta=0$ is 
$
S_0(U)=\frac{\d}{\d\theta} \log q_\theta(U)\Bigr|_{\theta=0}=\bigl(z-\E(Z)\bigr)^t \frac{\nabla p_X(U)}{p_X(U)}
$ 
where $\frac{\nabla p_X}{p_X}$ is the (nonparametric) score of $X$. 
Therefore, 
$
J_0(U)=\Var\bigl(S_0(U)\bigr)=\bigl(z-\E(Z)\bigr)^t \J(X)\bigl(z-\E(Z)\bigr)
$.
Plugging this expression into~\eqref{kullback2} gives~\eqref{debruijnexpansion} as required.
\end{proof}

In exploiting the parallelism between~\eqref{debruijnexpansion} and~\eqref{kullback}, this proof explains the presence of the $1/2$ factor in de Bruijn's identity: this is merely a second-order Taylor expansion factor due to the definition of Fisher information as the second derivative of divergence.
Besides, it is mentioned in~\cite{DemboCoverThomas91} that~\eqref{debruijnstandard} holds for any random vector $Z$ whose first four moments coincide with those of the standard Gaussian; here we see that it is sufficient that this condition hold for the second centered moments ($\Cov(Z)=\I$). 
 Also note that it is not required that $Z$ have a density. Thus,~\eqref{debruijn} also holds for a discrete valued perturbation $Z$.

\subsubsection{The Gaussian Case}

When $Z$ is Gaussian, de Bruijn's identity~\eqref{debruijn} is readily extended to positive values of~$t$. Simply substitute $X+\sqrt{t'}\,Z'$ for $X$, where $Z'$ is independent of $Z$ with the same distribution. By the stability property of the Gaussian distribution under convolution, $X+\sqrt{t'}\,Z'+\sqrt{t}\,Z$ and $X+\sqrt{t+t'}\,Z$ are identically distributed, and, therefore,
\begin{subequations}\label{debruijngauss}
\begin{equation}\label{debruijngaussgeneral}
\frac{d}{dt} h(X+\sqrt{t}\, Z) = \frac{1}{2}\,\tr\bigl(\J(X+\sqrt{t}\, Z)\,\Cov(Z)\bigr).
\end{equation}
For white $Z$, this reduces to 
\begin{equation}\label{debruijngausswhite}
\frac{d}{dt} h(X+\sqrt{t}\, Z)  = \frac{1}{2}\sigma^2_Z \, J(X+\sqrt{t}\, Z).
\end{equation}
\end{subequations}
Such a generalization cannot be established for non-Gaussian $Z$, because the Gaussian distribution is the only stable distribution with finite covariances.
Using the complementary relation~\eqref{fimmse} of Proposition~\ref{prop-fimmse} and making the change of variable $t'=1/t$, it is a simple matter of algebra Êto show that
\eqref{debruijngaussgeneral} is equivalent to
\begin{subequations}\label{debruijnmmse}
\begin{equation}\label{debruijngaussmmsegeneral}
\frac{d}{dt} h(\sqrt{t}\,X+ Z) = \frac{1}{2}\,\tr\bigl(\Cov(Z)^{-1}\,\Cov(X|\sqrt{t}\,X+ Z)\bigr).
\end{equation}
Since $\Cov(Z)^{-1}=\J(Z)$, this alternative identity also generalizes~\eqref{debruijngeneral} (with $X$ and $Z$ interchanged). For white $Z$, it reduces to
\begin{equation}\label{debruijngaussmmsewhite}
\frac{d}{dt} h(\sqrt{t}\,X+ Z) = \frac{1}{2\sigma^2_Z}\,\Var(X|\sqrt{t}\,X+ Z).
\end{equation}
\end{subequations}
The latter two identites were thoroughly investigated by Guo, Shamai and Verd\'u\cite{GuoShamaiVerdu05}. The above proof, via de Bruijn's identity and Kullback's expansion~\eqref{kullback}, is shorter than the proofs given in\cite{GuoShamaiVerdu05}, and also has an intuitive interpretation, as shown next.

\subsubsection{Intuitive Interpretations}

\begin{figure}[!ht]
\centering
\setlength{\unitlength}{0.06cm}
\begin{picture}(100,70)
\qbezier(0,40)(50,-40)(100,40)
\put(10,0){\vector(1,0){80}}
\put(50,0){\vector(0,1){50}}
\put(50,-5){\makebox(0,0){$\theta$}}
\put(50,60){\makebox(0,0){$D_X(p_{\theta}\|p_{\theta'})$}}
\put(95,0){\makebox(0,0){$\theta'$}}
%%\put(50,20){\vector(1,0){32}}\put(50,20){\vector(-1,0){32}}
\put(15,0){\dashbox{1}(70,20){}}
%\end{picture}
%\begin{picture}(100,100)
\qbezier(40,40)(50,-40)(60,40)
%\put(30,0){\vector(1,0){40}}
%\put(50,0){\vector(0,1){100}}
%\put(50,-5){\makebox(0,0){$\theta$}}
%\put(85,90){\makebox(0,0){$D_\X(p_{\theta}\|p_{\theta'})$}}
%\put(75,0){\makebox(0,0){$\theta'$}}
%%\put(50,20){\vector(1,0){5}}\put(50,20){\vector(-1,0){5}}
\put(43,0){\dashbox{1}(14,20){}}
\put(65,35){\makebox(0,0){\scriptsize(b)}}
\put(90,35){\makebox(0,0){\scriptsize(a)}}
\end{picture}
%\\[1cm]
%\makebox[\textwidth]{\hfill (a)~low value of $J_\theta(\X)$
%\hfill (b)~high value of $J_\theta(\X)$\hfill}
\caption{Kullback-Leibler divergence drawn as a function of the estimated parameter for (a)~low and (b)~high value of Fisher information.}\label{figfisher}
\end{figure}
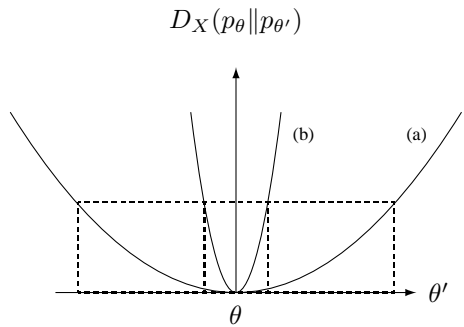

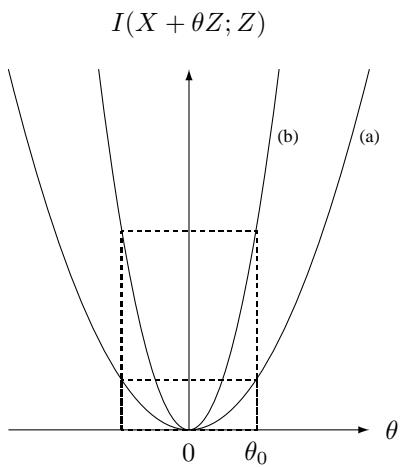
\begin{figure}[!ht]
\centering
\setlength{\unitlength}{0.06cm}
\begin{picture}(100,100)
\qbezier(10,80)(50,-80)(90,80)
\put(10,0){\vector(1,0){80}}
\put(50,0){\vector(0,1){80}}
\put(50,-5){\makebox(0,0){$0$}}
\put(65,-5){\makebox(0,0){$\theta_0$}}
\put(50,90){\makebox(0,0){$I(X+\theta Z;Z)$}}
\put(95,0){\makebox(0,0){$\theta$}}
\qbezier(30,80)(50,-80)(70,80)
\put(35,0){\dashbox{1}(30,44){}}
\put(35,0){\dashbox{1}(30,11){}}
\put(72,65){\makebox(0,0){\scriptsize(b)}}
\put(90,65){\makebox(0,0){\scriptsize(a)}}
\end{picture}
\caption{Mutual information between a noisy variable and the noise, drawn as a function of noise amplitude $\theta$ for (a)~low and (b)~high value of the variable's Fisher information.}\label{figfisher2}
\end{figure}

Expansions~\eqref{debruijnexpansion} and~\eqref{kullback} can be given similar interpretations. In~\eqref{kullback}, $D_X(p_\theta\|p_{\theta'})$
has local parabolic behavior at vertex $\theta=\theta'$ with curvature $=J_{\theta}(X)$, which means that for a given (small) value of divergence, $\theta$ is known all the more precisely as Fisher information $J_{\theta}(X)$ is large (see Fig.~\ref{figfisher}). This confirms that $J_{\theta}(X)$ is a quantity of ``information'' about $\theta$.
Similarly, \eqref{debruijnexpansion} shows that the mutual information $I(X+\theta Z; Z)$ between the noisy version $X+\theta\,Z$ of $X$ and the noise $Z$, seen as a function of the noise amplitude, is locally parabolic about $\theta=0$ with curvature $=J(X)$.
Hence for a given (small) value of noise amplitude $\theta_0$, the noisy variable is all the more dependent on the noise as $J(X)$ is higher (see Fig.~\ref{figfisher2}). 
Therefore, de Bruijn's identity merely states that Fisher information measures the \emph{sensitivity} to an arbitrary additive independent noise, in the sense that a highly ``sensitive'' variable, perturbed by a small additive noise, becomes rapidly noise-dependent as the amplitude of the noise increases. This measure of sensitivity of $X$ depends the noise covariances but is independent of the shape of the noise distribution otherwise, due to the fact that de Bruijn's identity remains true for non-Gaussian $Z$. Also, by the Cram\'er-Rao inequality~\eqref{cramerrao}, a Gaussian variable $X^*$ has lowest sensitivity to an arbitrary additive noise~$Z$. Thus the saddlepoint property of mutual information
$I(X+Z;Z) \geq I(X^*+Z;Z)$, classically established for Gaussian~$Z$~\cite{BordenMasonMcEliece85,DiggaviCover01,CoverThomas06} (see also Proposition~\ref{prop-saddle} below), is seen to hold to the first order of $\sigma^2_Z$ for an arbitrary additive noise $Z$.

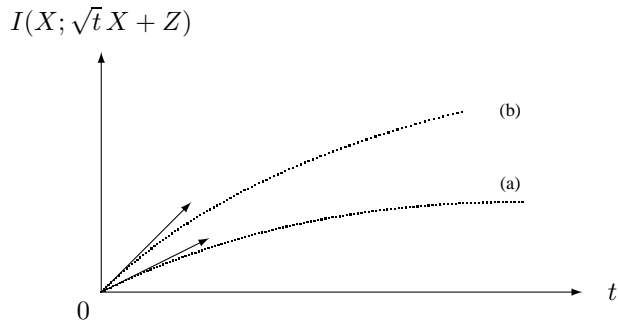
\begin{figure}[!ht]
\centering
\setlength{\unitlength}{0.08cm}
\begin{picture}(80,50)
\put(0,0){\vector(0,1){40}}
\put(0,0){\vector(1,0){80}}
\put(-3,-3){\makebox(0,0){$0$}}
\put(0,45){\makebox(0,0){$I(X;\sqrt{t}\,X+ Z)$}}
\put(85,0){\makebox(0,0){$t$}}
\qbezier[100](0,0)(20,20)(60,30)
\qbezier[120](0,0)(30,15)(70,15)
\put(0,0){\vector(1,1){15}}
\put(0,0){\vector(2,1){18}}
\put(68,18){\makebox(0,0){\scriptsize(a)}}
\put(68,30){\makebox(0,0){\scriptsize(b)}}
\end{picture}
\caption{Input-output mutual information over an additive noise channel, drawn as a function of SNR for small SNR and standard $Z$. (a)~Gaussian channel $J(Z)=1$. (b)~Laplacian channel $J(Z)=2$.}\label{figfisher3}
\end{figure}

A dual interpretation is obtained by exchanging the roles of $X$ and $Z$ in \eqref{debruijngeneral} or~\eqref{debruijnexpansion} to obtain an asymptotic formula for the input-output mutual information $I(X;\sqrt{t}\, X+Z)$ in a (non-Gaussian) additive noise channel $X\mapsto \sqrt{t}\,X+Z$ for small signal-to-noise ratio (SNR). In particular, for i.i.d. input entries or if the channel is memoryless, either $\Cov(X)$ or $\J(Z)$ is proportional to the identity matrix and, therefore,
\begin{equation}\label{debruijnprelov}
I(X;\sqrt{t}\, X+Z)= \frac{1}{2} J(Z) \, \sigma^2_X t +o(t)
\end{equation}
Thus, as has been observed in, e.g., \cite{LapidothShamai02, GuoShamaiVerdu05, GuoShamaiVerdu05a}, the rate of increase of mutual information per unit SNR is equal to $\frac{1}{2}J(Z)$ in the vicinity of zero SNR, regardless of the shape of the input distribution (see Fig.\ref{figfisher3}). 
In the case of a memoryless channel, it is also insensitive to input memory, since  in this case~\eqref{debruijnprelov} still holds for correlated inputs. Again by the Cram\'er-Rao inequality~\eqref{cramerrao}, the Gaussian channel exhibits a minimal rate of increase of mutual information, which complies with the well-known fact that non-Gaussian additive noise channels cannot have smaller capacity than that of the Gaussian channel.

\subsubsection{Applications}

Apart from its role in proving the EPI, de Bruijn's identity (Proposition~\ref{prop-debruijn}) has found many applications in the literature, although they were not always recognized as such. The Taylor expansion for non-Gaussianness corresponding to~\eqref{debruijnexpansion} in the scalar case ($n=1$) is mentioned, albeit in a disguised form, by Linnik\cite{Linnik59} who used it to prove the central limit theorem. Itoh\cite{Itoh70} used Linnik's expansion to characterize the Gaussian distribution by rotation. 
Similar expansions have been derived by Prelov and others (see, e.g., \cite{Prelov70,Prelov72,IbragimovKhasminskii72,Prelov88, Prelov89,PrelovMeulen93,PinskerPrelovVerdu95,PinskerPrelovMeulen97, PinskerPrelovMeulen98, PinskerPrelovMeulen98a,PrelovMeulen03,PrelovMeulen03a,PrelovVerdu04}) to investigate the behavior of the capacity or mutual information in additive Gaussian or non-Gaussian noise channels under various asymptotic scenarios. In particular, \eqref{debruijnprelov} was apparently first stated explicitly by Pinsker, Prelov and van der Meulen~\cite{PinskerPrelovMeulen98}. A similar result was previously published by Verd\'u~\cite{Verdu90} (see also \cite{Verdu02}) who used
Kullback's expansion~\eqref{kullback} to lower bound the capacity per unit SNR for non-Gaussian memoryless additive noise channels, a result which is also an easy consequence of~\eqref{debruijnprelov}. Motivated by the blind source separation problem, Pham~\cite{Pham05} (see also~\cite{PhamVrins05,VrinsPhamVerleysen07}) investigated the first and second-order expansions in $\theta$ of entropy for non-Gaussian perturbation $Z$ (not necessarily independent of $X$) and recovers de Bruijn's identity as a special case. Similar first and second-order expansions for mutual information in non-Gaussian additive noise channels were derived by Guo, Shamai and Verd\'u~\cite{GuoShamaiVerdu05a}, yielding~\eqref{debruijnprelov} as a special case.

\subsubsection{Generalized De Bruijn's Identity}

Palomar and Verd\'u~\cite{PalomarVerdu06} proposed a matrix version of de Bruijn's identity by considering the gradient of $h(X+Z)$ with respect to the noise covariance matrix $\Cov(Z)$ for Gaussian $Z$. We call attention that this is a simple consequence of~\eqref{debruijngeneral}; the generalization to non-Gaussian $Z$ is as follows. 
\begin{corollary} \label{cor-debruijn}
%Under the same assumption as in Proposition~\ref{prop-debruijn},
\begin{equation}\label{debruijnpalomarverdu}
\frac{d}{d\K}h(X+Z)\Bigr|_{\K=\0} = \frac{1}{2} \J(X),
\end{equation}
where we have noted $\K=\Cov(Z)$.
\end{corollary}
\begin{proof}[Proof\/\footnote{The $1/2$ factor is absent in~\cite{PalomarVerdu06}, due to the fact that complex gradients are considered.}]
By~\eqref{debruijngeneral}, we have the following expansion:
$$
h(X+Z) = \frac{1}{2}\,\tr\bigl(\J(X)\,\K\bigr) + o(\|\K\|)
$$h
where $\|\K\|$ denotes the Fr\"obenius norm  of $\K=\K^t$. But this is of the form of a first-order Taylor expansion of a function with respect to a matrix\footnote{Putting the matrix entries into a column vector $\mathbf{k}$ it is easily found that $\tr \bigl(\frac{df}{d\K}(\0)\cdot \K^t \bigr)=\k^t\frac{df}{d\k}(\0)$. }:
$$
f(\K)=f(\bm{0})+ \tr \bigl(\frac{df}{d\K}(\bm{0})\cdot \K^t \bigr) + o(\|\K\|),
$$
and~\eqref{debruijnpalomarverdu} follows by identifiying the gradient matrix.
\end{proof}

\subsubsection{Relationship between the Cram\'er-Rao Inequality and a Saddlepoint Property of Mutual Information}\label{sec-saddle}
The following saddle point property of mutual information, which  was proved in\cite{Ihara78} using a result of Pinsker\cite{Pinsker56}, states that the worst possible noise distribution in a additive noise channel is the Gaussian distribution.
\begin{proposition}\label{prop-saddle}
Let $X$ be any random vector, and let $X^*$ be a Gaussian random vector with identical second moments. For any Gaussian random vector $Z$ independent of $X$ and $X^*$,
\begin{equation}\label{saddle}
I(X+Z;Z) \geq I(X^*+Z;Z).
\end{equation}
\end{proposition}
\medskip
\begin{proof}[Proof (following\cite{DiggaviCover01})]
Noting that $Y^*=X^*+Z$ has identical second moments as $Y=X+Z$, we have
$I(X+Z;Z) - I(X^*+Z;Z) = h(Y)-h(X)-h(Y^*)+h(X^*)=D(X\|X^*)-D(Y\|Y^*)$. The result follows by the data processing inequality for divergence, applied to the transformation $X\to Y=X+Z$.
\end{proof}
This proof, in constrast to that given in~\cite{BordenMasonMcEliece85 ,CoverThomas06} for scalar variables, does not require the EPI, and is through a much less involved argument.

Interestingly, by virtue of de Bruijn's identity, it can be shown that~\eqref{saddle} is equivalent to the famous Cram\'er-Rao inequality\footnote{This follows from the relation $\J(X)-\J(X^*)=\Cov\bigl(S(X)-S^*(X)\bigr)\geq 0$, where $S^*(X)$ is defined as in~\eqref{nongaussJ}.}
\begin{equation}\label{crb}
\J(X)\geq \J(X^*)=\Cov(X)^{-1}.
\end{equation}
To see this, divide both sides of \eqref{saddle} by the entries of $\Cov(Z)$ and let $\Cov(Z)\to\0$. By Corollary~\ref{cor-debruijn}, this gives $\frac{1}{2} \J(X) \geq \frac{1}{2} \J(X^*)$. Conversely, integrating the relation $\frac{1}{2} \tr\bigl(\J(X+Z)\Cov(Z)\bigr) \geq \frac{1}{2} \tr\bigl(\J(X^*+Z)\Cov(Z)\bigr)$ using de Bruijn's identity~\eqref{debruijngauss} readily gives \eqref{saddle}.

\subsection{Earlier Proofs of the EPI}\label{sec-fiiepi}

All available information theoretic proofs of the EPI use de Bruijn's identity to integrate the FII (or the corresponding inequality for MMSE) over the path of a continuous Gaussian perturbation. 
To simplify the presentation, we first consider a path of the form $\{X+\sqrt{t}\, Z\}_{t\in (0;+\infty[}$ where $Z$ is assumed \emph{standard} Gaussian. The derivations in this section are readily extended to the case where $Z$ is arbitrary Gaussian, by means of the corresponding generalized FII and de Bruijn's identity.

\subsubsection{Basic Proof} \label{stamepiproof}
The following is a simplified version of Stam's proof~\cite{Stam59}. Apply the FII~\eqref{fii3} to the random vectors $(X_i+\sqrt{t}\, Z_i)_i$, where the $(Z_i)_i$ are independent and standard Gaussian. This gives $J(\sum_i a_i X_i +\sqrt{t}\, Z) - \sum_i a_i^2 J(X_i+\sqrt{t}\,Z_i)\leq 0$, where $Z=\sum_i a_i Z_i$ is also standard Gaussian.
By de Bruijn's identity~\eqref{debruijngausswhite}, it follows that
$
f(t)=h(\sum_i a_i X_i +\sqrt{t}\, Z) - \sum_i a_i^2 h(X_i+\sqrt{t}\,Z_i)
$
is a nonincreasing function of $t$. But $f(t)=h(t^{-1/2}\sum_i a_i X_i + Z) - \sum_i a_i^2 h(t^{-1/2}X_i+Z_i)$ tends to $h(Z) - \sum_i a_i^2 h(Z_i) =0$ as $t\to\infty$ (see Lemma~\ref{lem-continuity} below). Therefore, $f(0)\geq f(\infty)=0$, which is the EPI~\eqref{epi3}.

Note that the case of equality in~\eqref{epi3} is easily determined by this approach, since it reduces to the case of equality in the corresponding FII (see Section~\ref{sec-eqfii}). Namely, equality holds in the EPI~\eqref{epi3} iff all random vectors $X_i$ for which $a_i\ne 0$ are Gaussian with \emph{identical} covariances. It follows that equality holds in the classical form of the EPI~\eqref{epi1} iff all random vectors $X_i$ for which $a_i\ne 0$ are Gaussian with \emph{proportional} covariances.

\subsubsection{Integral Representations of Differential Entropy}
In the above proof, de Bruijn's identity can be rewritten as an integral representation of entropy. To see this, introduce an auxiliary Gaussian random vector $X^*$, and rewrite de Bruijn identity~\eqref{debruijngausswhite} in the form\footnote{When $X^*$ is chosen such that $\Cov(X^*)=\Cov(X)$, the identity relates nonnegative ``non-Gaussiannesses''~\eqref{nongaussH} and~\eqref{nongaussJ}.} $\frac{d}{dt} \bigl(h(X^*+\sqrt{t}\, Z)-h(X+\sqrt{t}\, Z)\bigr) = -  \frac{1}{2} \bigl(J(X+\sqrt{t}\, Z)-J(X^*+\sqrt{t}\, Z)\bigr)$. Since $h(X^*+\sqrt{t}\, Z)-h(X+\sqrt{t}\, Z)\to 0$ as $t\to\infty$, we may integrate from $t=0$ to $+\infty$ to obtain $h(X)-h(X^*)$ as the integral of $J(X+\sqrt{t}\, Z)-J(X^*+\sqrt{t}\, Z)$. If, for example, $Z^*$ is chosen standard, one obtains the integral representation~\cite{MadimanBarron06}
\begin{subequations}\label{integrals}
\begin{equation}\label{i1}
h(X) - \frac{n}{2}\log(2\pi e) = - \frac{1}{2}\int_0^\infty J(X+\sqrt{t}\,Z)-\frac{n}{1+t} \, dt
\end{equation}
In view of this identity, the EPI~\eqref{epi3} immediately follows from the corresponding FII~\eqref{fii3}.

\subsubsection{Other Paths of Integration} \label{sec-paths}
Several variants of the above proof were published, either in differential or integral form. Dembo, Cover and Thomas~\cite{DemboCoverThomas91} and Carlen and Soffer~\cite{CarlenSoffer91} use a path connecting $Z$ to $X$ of the form $\{\sqrt{t}\, X+\sqrt{1-t}\, Z\}_{t\in (0;1)}$. The argument leading to the EPI is the same up to an appropriate change of variable. The corresponding integral representation
\begin{equation}\label{i2}
h(X)=\frac{n}{2}\log(2\pi e) - \frac{1}{2}\int_0^1 J(\sqrt{t}\,X+\sqrt{1-t}\,Z)-n \, \frac{dt}{t}
\end{equation}
was first used by Barron~\cite{Barron86} to prove a strong version of the central limit theorem.
Verd\'u and Guo~\cite{VerduGuo06} used the path $\{\sqrt{t}\,X+ Z\}_{t\in (0;+\infty[}$ and replaced Fisher information by MMSE. They used~\eqref{debruijngaussmmsewhite} to integrate inequality~\eqref{mmsei} over this path. Their proof is completely equivalent to Stam's proof above, by means of the complementary relation~\eqref{fimmsewhite} of Proposition~\ref{prop-fimmse} and the change of variable $t'=1/t$. The corresponding integral representation becomes~\cite{GuoShamaiVerdu05,VerduGuo06,GuoShamaiVerdu06a}
\begin{equation}\label{i3}
h(X)=\frac{n}{2}\log(2\pi e) - \frac{1}{2}\int_0^\infty \frac{n}{1+t} - \Var(X|\sqrt{t}\,X+Z) \, dt.
\end{equation}
Yet another possibility is to take the path $\{\sqrt{1-t}\, X+\sqrt{t}\, Z\}_{t\in (0;1)}$ connecting $X$ to $Z$, leading to the following integral representation:
\begin{equation}
h(X)=\frac{n}{2}\log(2\pi e) - \frac{1}{2}\int_0^\infty n - \frac{1}{t}\Var(X|\sqrt{1-t}\,X+\sqrt{t}\,Z) \, \frac{dt}{t}.
\end{equation}
\end{subequations}
All the above representations for entropy are equivalent through appropriate changes of variable inside the integrals.
 
\section{A New Proof of Shannon's EPI} \label{sec-epi}
 
\subsection{A Mutual Information Inequality (MII)}

From the analysis made in Section~\ref{sec-earlier}, it is clear that earlier information theoretic proofs of the EPI can be seen as variants of the same proof, with the following common ingredients:
\begin{enumerate}
\item a \emph{data processing inequality} applied to the linear transformation~\eqref{transformation}.
\item an integration over a path of a continuous  \emph{Gaussian perturbation}.
\end{enumerate}
While step~1) uses the data processing theorem in terms of either parametric Fisher information or MMSE, step 2) uses de Bruijn's identity, which relates Fisher information or MMSE to entropy or mutual information.  
This suggests that it should be possible to prove the EPI via a data processing argument made directly on the mutual information. The interest is two-fold: First, compared to the data processing theorem for Fisher information, the corresponding theorem for Shannon's mutual information is presumably more familiar to the readers of this journal. Second, this approach sidesteps both Fisher information and MMSE and avoids the use of de Bruijn's identity~\eqref{debruijngausswhite} or~\eqref{debruijngaussmmsewhite}.

We shall prove a stronger statement than the EPI, namely, that the difference between both sides of~\eqref{epi3} decreases as independent Gaussian noise $Z$ is added. Since $h(X+Z)-h(X)=I(X+Z;Z)$ for any $X$ independent of $Z$ (see Lemma~\ref{lem-entropy} below), we write this statement in terms of mutual information as follows.

\begin{theorem}[Mutual Information Inequality (MII)]\label{th-mii}
For finitely many independent random $n$-vectors $(X_i)_i$ with finite covariances, any real-valued coefficients $(a_i)_i$ normalized such that $\sum_i a_i^2 = 1$, and any Gaussian $n$-vector $Z$ independent of $(X_i)_i$,
\begin{equation}\label{mii}
I(\sum_i a_i X_i + Z;Z) \leq \sum_i a_i^2 I(X_i+Z;Z).
\end{equation}
Furthermore, this inequality implies the EPI~\eqref{epi3}.
\end{theorem}

The MII~\eqref{mii} can be interpreted as a convexity property of mutual information  under the covariance-preserving transformation~\eqref{transformation}.
As we shall see, the crucial step in the proof of Theorem~\ref{th-mii} is the data processing inequality for mutual information~\cite{CoverThomas06}. We also need the following technical lemmas. In order to  be mathematically correct throughout we first establish some basic properties of mutual information and entropy.

\begin{lemma}\label{lem-entropy}
Let $X$ be any random $n$-vector and $Z$ be any Gaussian $n$-vector independent of $X$. Then $X+Z$ has a density, $h(X+Z)$ exists and is finite. In addition, if $h(X)$ exists, %as in Proposition~\ref{prop-hdef}, 
the identity $I(X+Z;Z)=h(X+Z)-h(X)$ always holds.
\end{lemma}

\begin{proof}
Let $\phi_{X}(u)=\E\bigl(\exp(ju\cdot X)\bigr)$ be the characteristic function of $X$; that of $Y=X+Z$ is $\phi_{Y}(u)=\phi_{X}(u)\phi_{Z}(u)$ where $\phi_{Z}(u)= \exp(ju\cdot\E(Z)-\frac{1}{2}u^{t}\Cov(Z)u)$. Since characteristic functions are bounded continuous and  $\phi_{Z}(u) $ has rapid decay (faster than any inverse of a polynomial) at infinity, it follows that $\phi_{Y}(u)$ is integrable. Therefore, $Y$ admits a bounded density\footnote{This density is in fact indefinitely differentiable and strictly positive, and all its derivatives are bounded and tend to zero at infinity.} $p(y)$, such that $p(y)\leq c$ where $c$ is some positive constant. The negative part of the integral $-\int p(y) \log p(y) dy$ is
$h^{-}(Y)=\int_{p(y)\geq 1} p(y)\log p(y) dy \leq \log c$, which is bounded.
Hence $h(X+Z)=h(Y)$ exists and is finite.

If $h(X)$ exists, then either $X$ admits a density $p_{X}(x)$ or it does not. In the former case $(Y,Z)=(X+Z,Z)$ also admits a density $p(y,z)=p_{X}(y-z)p_{Z}(z)$ and the identity $I(X+Z;Z)=h(X+Z)-h(X)$ is well known. In the latter case we have put $h(X)=-\infty$ (see Section~\ref{sec-hdef}). Since $X$ is not absolutely continuous with respect to the Lebesgue measure, there exists a set $A$ of zero measure such that $P(X\in A)>0$. Then $B=\{(y,z) \mid y-z\in A\}$ has zero Lebesgue measure and $P\bigl((X+Z,Z)\in B\bigr)=P(X\in A)>0$. Since $B$ is also of zero measure with respect to the product probability measure with density $p_{Y}(y)p_{Z}(z)$, it follows that $(Y,Z)$ is not absolutely continuous with respect to this product measure. Therefore, by the theorem of Gel'fand-Yaglom-Perez~\cite[chap.~2]{Pinsker64}, one has $I(X+Z;Z)=I(Y;Z)=+\infty$ and the identity $I(X+Z;Z)=h(X+Z)-h(X)$ still holds.
\end{proof}
In the same way one can prove that the identity $I(X+\sqrt{t}\,Z;Z)=h(X+\sqrt{t}\,Z)-h(X)$ always holds for any $t\geq 0$.

The following inequality~\eqref{sato} was proved for two variables by Sato~\cite{Sato78} who used it to derive an outer bound to the capacity region of broadcast channels. A similar inequality appears in\cite[Thm. 4.2.1]{Gallager68} and in\cite[Thm 1.9]{McEliece02}.

\begin{lemma}[Sato's Inequality]\label{lem-sato}
If the random vectors $(X_i)_i$ are independent of $Z$ and of each other, then
\begin{equation}\label{sato}
I\bigl((X_i+Z)_i;Z\bigr) \leq \sum_i I(X_i+Z;Z).
\end{equation}
\end{lemma} 

\begin{proof}
Let $Y_i=X_i+Z$ for all $i$. By the chain rule for mutual information~\cite[chap.~3]{Pinsker64}, one has
\begin{subequations}
\begin{align}
I\bigl((Y_i)_i;Z\bigr) &=
\sum_{i} I(Y_{i};Z | Y_{1},\ldots,Y_{i-1})\\
&=\sum_{i} I(Y_{i};Z, Y_{1},\ldots,Y_{i-1}) - I(Y_{i};Y_{1},\ldots,Y_{i-1})\\
&\leq \sum_{i} I(Y_{i};Z, Y_{1},\ldots,Y_{i-1}) \\
&=\sum_{i} I(Y_{i};Z) - I(Y_{i};Y_{1},\ldots,Y_{i-1}|Z) = \sum_{i}I(Y_{i};Z) .
\end{align}
\end{subequations}
\end{proof}

An alternative proof in the case where $\Y=(Y_i)_i$ admits a density, is as follows. Define the \emph{symmetric mutual information} between the components of $\Y$ by the divergence
\begin{equation}\label{symmetricmi}
I\{(Y_i)_i\}=\E \log \frac{p(\Y)}{\prod_i p(Y_i)} = \sum_{i>1} I(Y_i;Y_1,\ldots,Y_{i-1})
\end{equation}
From the definitions it is obvious that
$I\bigl((Y_i)_i;Z\bigr) - \sum_i I(Y_i;Z) = I\{(Y_i)_i|Z\} - I\{(Y_i)_i\}$.
The result follows since $I\{(Y_i)_i\}\geq 0$ and  $I\{(Y_i)_i|Z\}=I\{(X_i)_i\}=0$.

\begin{lemma}\label{lem-continuity}
If $X$ and $Z$ are independent random $n$-vectors with finite covariances and differential entropies, then 
\begin{equation}\label{continuity}
\lim_{t\to 0^{+}} I(X+\sqrt{t} Z;Z) = 0.
\end{equation}
If, in addition, $I(X+\sqrt{t}Z;Z)$ is differentiable at $t=0$, then 
\begin{equation}\label{mitaylor}
I(X+a\sqrt{t} Z; Z) = a^2 I(X+ \sqrt{t} Z; Z) + o(t)
\end{equation}
where $o(t)$ is a function defined for all $t\geq0$ such that $o(t)/t\to 0$ as $t\to 0^{+}$.
\end{lemma}

\begin{proof}
To prove~\eqref{continuity}, let $X_{t}=X+\sqrt{t}\,Z$. Taking characteristic functions, $\phi_{X_{t}}(u)=\phi_{X}(u)\phi_{Z}(\sqrt{t}u)\to  \phi_{X}(u)$ as $t\to 0^{+}$. Therefore $X_{t}\to X$ %in probability and therefore
 in distribution. Let $X^*$ and $Z^*$ be Gaussian $n$-vectors have identical covariances as $X$ and $Z$, respectively. Likewise $X^{*}_{t}=X^{*}+\sqrt{t}Z^{*}\to X^{*}$ in distribution. By the lower semi-continuity of divergence (see \cite[\S~Ê2.4]{Pinsker64} and~\cite[Thm.~Ê1]{Posner75}), we have the inequality 
\begin{equation}
D(X\|X^*)\leq \liminf_{t\to 0^{+}} D(X_{t}\| X_{t}^*). \label{eq-lowersc}
\end{equation}
The following quantities are all finite.
\begin{subequations}
\begin{align}
D(X_{t}\| X_{t}^*)-D(X\|X^{*})&=h(X_{t}^{*})-h(X_{t})-h(X^{*})+h(X^{*})\\
&=h(X^{*}+\sqrt{t}Z^{*})-h(X^{*}) - I(X+\sqrt{t}Z;Z)
\end{align}
\end{subequations}
An easy calculation for Gaussian vectors gives $\lim_{t\to 0^{+}}h(X^{*}+\sqrt{t}Z^{*})=h(X^{*})$. Therefore~\eqref{eq-lowersc} reduces to
\begin{equation}
\limsup_{t\to 0^{+}} I(X+\sqrt{t}Z;Z) \leq \liminf_{t\to 0^{+}}h(X^{*}+\sqrt{t}Z^{*}) - h(X^{*})=0.
\end{equation}
This combined with nonnegativity of mutual information proves~\eqref{continuity}.

Now suppose $I(X+\sqrt{t}Z;Z)$ is differentiable at $t=0$. 
Since $\lim_{t\to 0^{+}}I(X+\sqrt{t}\,Z;Z)=0$, for any $a\in\R$, 
$I(X+a\sqrt{t} Z; Z)/a^{2}t$ and $I(X+ \sqrt{t} Z; Z)/t$ tend toward the same limit as $t\to 0^{+}$. This reduces to~\eqref{mitaylor}.
\end{proof}

Note that neither Lemma~\ref{lem-sato} nor Lemma~\ref{lem-continuity} requires $Z$ to be Gaussian. 
The following lemma gives an important situation where the differentiability assumption of Lemma~\ref{lem-continuity} is met.

\begin{lemma}\label{lem-gausscase}
Let $X$ be any random $n$-vector with finite covariances and differential entropy, and let $Z,Z'$ be identically distributed Gaussian $n$-vectors such that $X,Z,Z'$ are independent. The quantity $I(X+\sqrt{t}Z;Z)$ is differentiable at any $t>0$.
In addition, if $X'=X+\sqrt{u}\,Z'$ where $u>0$, then %for all $t\geq 0$,
\begin{equation}\label{i}
I(X'+\sqrt{t}Z;Z) = I(X+\sqrt{u+t}\,Z;Z)-I(X+\sqrt{u}\, Z; Z) 
\end{equation}
is also differentiable at $t=0$.
\end{lemma}

\begin{proof}
Following Stam~\cite{Stam59}, Barron~\cite{Barron86} proved that $h(X+\sqrt{t}Z)$ is differentiable in $t>0$ for any square-integrable~$X$. The proof involves exchanges of differentiation and expectation justified by the dominated convergence theorem and is not repeated here. From Lemma~\ref{lem-entropy} is follows that $I(X+\sqrt{t}Z;Z)$ is likewise differentiable at any $t>0$.
Now the following quantities are all finite.
\begin{subequations}\label{midiff}
\begin{align}
I(X'+\sqrt{t}Z;Z)&= I(X+\sqrt{u}\, Z'+\sqrt{t}\,Z; Z)\\
&=h(X+\sqrt{u}\, Z'+\sqrt{t}\,Z)-h(X+\sqrt{u}\, Z')\\
&=h(X+\sqrt{u}\, Z'+\sqrt{t}\,Z)-h(X) + h(X)-h(X+\sqrt{u}\, Z')\\
&=I(X+\sqrt{u}\, Z'+\sqrt{t}\,Z;\sqrt{u}\, Z'+\sqrt{t}\,Z) - I(X+\sqrt{u}\, Z';\sqrt{u}\, Z')\\
&=I(X+\sqrt{u+t}\,Z;Z)-I(X+\sqrt{u}\, Z; Z) 
\end{align}
\end{subequations}
The last equality follows from the stability property of the Gaussian distribution under convolution, since $\sqrt{u}\, Z'+\sqrt{t}\,Z$ is identically distributed as $\sqrt{u+t}\, Z$. 
Since $I(X+\sqrt{u+t}\,Z;Z)$ is differentiable at $t=0$ for any $u>0$, $I(X'+\sqrt{t}\,Z;Z)=I(X+\sqrt{u+t}\,Z;Z)-I(X+\sqrt{u}\, Z; Z)$ is likewise differentiable at $t=0$. 
\end{proof}

\begin{proof}[Proof of Theorem~\ref{th-mii}]
We may always assume that $a_{i}\ne 0$ for all $i$---otherwise simple delete the $X_{i}$ for which $a_{i}=0$. To prove~\eqref{mii}, we may also assume that all the $X_{i}$ have finite differential entropies, since otherwise the right-hand side of~\eqref{mii} is $=+\infty$ by Lemma~\ref{lem-entropy}. Then all the $X_{i}$ admit densities, and $\sum_{i}a_{i}X_{i}$ likewise admits a density and has finite covariances. From Proposition~\ref{prop-hdef} it follows that $h(\sum_{i}a_{i}X_{i})<+\infty$, and since conditioning reduces entropy, $-\infty<h(X_{i})\leq h(\sum_{i}a_{i}X_{i})$. Therefore, $\sum_{i}a_{i}X_{i}$ also has finite differential entropy. From this and Lemma~\ref{lem-entropy} it follows that all subsequent mutual informations will be finite.

We can write the following string of inequalities: 
\begin{subequations}\label{a}
\begin{align}
I(\sum_i a_i X_i + Z;Z) &=I(\sum_i a_i (X_i + a_i Z);Z) \label{a1}\\
&\leq I( (X_i + a_i Z)_i ;Z)  \label{a2}\\
&\leq \sum_i I(X_i + a_i Z ;Z)\label{a3} %\\
\end{align}
\end{subequations}
where~\eqref{a1} holds since $\sum_ia_i^2=1$, \eqref{a2}~follows from the data processing theorem applied to the linear transformation~\eqref{transformation}, \eqref{a3}~follows from Sato's inequality (Lemma~\ref{lem-sato}).
Note that substituting $\sqrt{t}Z$ for $Z$ in~\eqref{a3} and
assuming that $Z$ and the $X_i$ satisfy the differentiability assumption of Lemma~\ref{lem-continuity} for all $i$, one obtains
\begin{equation}\label{a4}
I(\sum_i a_i X_i + \sqrt{t}Z;Z) \leq  \sum_i a_i^2 I(X_i+\sqrt{t}Z;Z) + o(t)
\end{equation}
We now use the assumption that $Z$ is \emph{Gaussian} to eliminate the $o(t)$ term in~\eqref{a4}.

Let $X'_i=X_i+\sqrt{u}\,Z'_i$ for all $i$ and $u>0$, where the 
$Z'_i$ are Gaussian, identically distributed as $Z$ but independent of all other random vectors. Then $Z'=\sum_i a_i Z'_i$ is identically distributed as $Z$, and applying~\eqref{a3} to the $X'_i$ and to $\sqrt{t}\,Z$, one obtains
\begin{subequations}\label{b}
\begin{align}
I(\sum_i a_i X_i+\sqrt{u}\,Z'+\sqrt{t}Z;Z) &\leq 
\sum_{i} I(X'_{i}+a_{i}Z;Z)\\
&=\sum_i a_i^2 I(X_i+\sqrt{u}\,Z'_i+\sqrt{t}Z;Z)+ o(t)\label{a5}
\end{align}
\end{subequations}
where the last equality follows from the fact that by Lemma~\ref{lem-gausscase}, 
 the $X'_{i}=X_i+\sqrt{u}\,Z'_i$ satisfy the differentiability assumption of Lemma~\ref{lem-continuity}. 
Now define
$$
f(t)=I(\sum_i a_i X_i+\sqrt{t}Z;Z) - \sum_i a_i^2 I(X_i+\sqrt{t}Z;Z).
$$
Using~\eqref{i}, inequality~\eqref{b} is easily rewritten as
$$
I(\sum_i a_i X_i +\sqrt{u+t}\, Z; Z)-I(\sum_i a_i X_i +\sqrt{u}\, Z; Z) \leq
\sum_i a_i^2  \bigl( I(X_i+\sqrt{u+t}\,Z;Z) - I(X_i+\sqrt{u}\,Z;Z)\bigr) + o(\eps),
$$
that is $f(u+t)\leq f(u)+o(t)$ for any $u>0$. Since $f(u)$ is differentiable at any $u>0$ by Lemma~\ref{lem-gausscase}, it easily follows that $f(u)$ is non-increasing in $u>0$. Also, by Lemma~\ref{lem-continuity}, $\lim_{t\to 0} f(t)= f(0)=0$.
Therefore, $f(1)\leq f(0)=0$, which is the required MII~\eqref{mii}.

Finally, we show that the MII implies the EPI~\eqref{epi3}. Since $\sum_i a_i^2=1$ and $I(X+Z;Z)=h(X+Z)-h(X)=I(X;X+Z)+h(Z)-h(X)$ for $X$ independent of $Z$, \eqref{mii} can be rewritten as
\begin{equation}\label{forverdu}
h(\sum_i a_i X_i ) - \sum_i a_i^2 h(X_i) \geq I(\sum_i a_i X_i; \sum_i a_i X_i + Z) - \sum_i a_i^2 I(X_i;X_i+Z).
\end{equation}
Now replace $Z$ by $\sqrt{t}\, Z$ and let $t\to \infty$. The terms in the right-hand side of the above inequality are of the form $I(X;X+\sqrt{t}\,Z)=I(X;\frac{1}{\sqrt{t}}\, X+Z)$, which tends to zero as $t\to\infty$ by Lemma~\ref{lem-continuity}. This completes the proof.
\end{proof}

\subsection{Insights and Discussions}

\subsubsection{Relationship to Earlier Proofs}
Of course, Theorem~\ref{th-mii} could also be proved using the conventional techniques of Section~\ref{sec-earlier}. In fact, it follows easily from
either one of the integral representations~\eqref{integrals}. Also Lemma~\ref{lem-continuity} is an easy consequence of de Bruijn's identity, since by~\eqref{debruijnexpansion}, both sides of~\eqref{mitaylor} are equal to $\frac{t}{2} \tr \bigl(\Cov(aZ)\J(X)\bigr)= \frac{a^2t}{2} \tr \bigl(\Cov(Z)\J(X)\bigr)$.
The originality here lies in the above proof of Theorem~\ref{th-mii} and the EPI, which in contrast to existing proofs, requires neither de Bruijn's identity nor the notions of Fisher information or MMSE.

The new proof shares common ingredients with earlier proofs of the EPI, namely items~1) and~2) listed at the beginning of this section. The difference is that they are used directly in terms of mutual information. As in section~\ref{sec-paths}, other paths of continuous Gaussian perturbation could very well be used, through suitable changes of variable. 

One may wonder if mutual informations in the form $I(X;\sqrt{t}\, X+Z)$ rather than $I(X+\sqrt{t}\, Z; Z)$ could be used in the above derivation of Theorem~\ref{th-mii}, particularly in inequalities~\eqref{a}. This would offer a dual proof, in the same way as Verd\'u and Guo's proof is dual to Stam and Blachman's original proof of the EPI, as explained in section~\ref{sec-earlier}. But a closer look at the above proof reveals that the dual approach would amount to prove~\eqref{forverdu}, whose natural proof using the data processing inequality is through~\eqref{a}. Thus, it turns out that the two approaches amount to the same.

Also note that by application of de Bruijn's identity, inequality~\eqref{a4} reduces to the FII~\eqref{fii3} . Thus the MII~\eqref{mii} implies both the EPI~\eqref{epi3} and the FII~\eqref{fii3}.

\subsubsection{The Equality Case}
Our method does not easily settle the case of equality in the MII. By the preceding remark, however, equality in~\eqref{a4} implies equality in the FII~\eqref{fii3}, which was determined in Section~\ref{sec-eqfii}. It follows that equality holds in the MII~\eqref{mii} if and only if all random vectors $X_i$ such that $a_i\ne 0$ are Gaussian with identical covariances. This result implies the corresponding necessity condition of equality in the EPI, but is not evident from the properties of mutual information alone. 

\subsubsection{On the Gaussianness of $Z$}
It is interesting to note that from~\eqref{a4}, the MII holds up to first order of the noise variance, regardless of whether $Z$ is Gaussian or not. 
However, the stability property of the Gaussian distribution under convolution was crucial in the next step of the proof, because the Gaussian perturbation $Z$ can be made to affect the random vectors independently. In fact, the MII can be easily rewritten as
\begin{equation}\label{mii3}
h(\sum_i a_i X_i ) - \sum_i a_i^2 h(X_i) \geq h(\sum_i a_i X'_i) - \sum_i a_i^2 h(X'_i)
\end{equation}
where $X'_i=X_i+Z_i$ for all $i$, the $(Z_i)_i$ being independent copies of $Z$.
This does \emph{not} hold in general for non-Gaussian random vectors $(Z_i)_i$. To see this, choose $(X_i)_i$ themselves Gaussian with identical covariances. Then the left-hand side of~\eqref{mii3} is zero, and by the necessity of the condition for equality in the EPI, the right-hand side is positive, as soon as $Z_i$ is non-Gaussian for some $i$ such that $a_i\ne 0$. Therefore, in this case, the opposite inequality is obtained. In other words, adding non-Gaussian noise may \emph{increase} the difference between both sides of the EPI~\eqref{epi3}, in accordance with the fact that this difference is zero for Gaussian random vectors.

\subsubsection{On the finite second-order moment assumption} To prove Theorem~\ref{th-mii} we have assumed for simplicity that the $X_{i}$ have finite covariances so that differential entropies are well-defined and the  lower semi-continuity argument in Lemma~\ref{lem-continuity} applies. However, it would be possible to weaken this condition to first-order finite moment or even to the condition of Proposition~\ref{prop-hdef} by considering divergences with respect to probability distributions other than Gaussian, e.g. exponential or Cauchy distributions as in the proof of Proposition~\ref{prop-hdef}.
The details are left to the reader.

\subsubsection{On the Use of Sato's Inequality}

Sato used~\eqref{sato} and the data processing inequality to derive his cooperative outer bound to the capacity region of two-user broadcast channels\cite{Sato78}. This bound was used to determine the capacity of a two-user MIMO Gaussian broadcast channel\cite{CaireShamai03}. Sato's bound was later replaced by the EPI to generalize Bergmans' solution to an arbitrary multi-user MIMO Gaussian broadcast channel using the notion of an ``enhanced'' channel~\cite{WeingartenSteinbergShamai06}. In the present paper, the EPI itself is proved using Sato's inequality and the data processing inequality.
This suggests that for proving converse coding theorems, a direct use of the EPI may be avoided by suitable inequalities for mutual information. A similar remark goes for the generalization of Ozarow's solution to vector Gaussian multiple descriptions\cite{WangViswanath07}.

\subsubsection{Relationship Between Various Data Processing Theorems}
\label{sec-dpimi}
Proposition~\ref{prop-dpi} enlightens the connection between two estimation theoretic data processing inequalities: parametric (Fisher information) and nonparametric (MMSE). While these were applied in earlier proofs of the EPI, the new proof uses the same data processing argument in terms of mutual information: any transformation $X\to Y$ in a Markov chain $\theta\to X\to Y$ reduces information about $\theta$. This can also be given a parametric form using divergence~\eqref{divergence}. Thus, if $\theta\to X\to Y$ form a Markov chain, then
\begin{subequations}
\begin{align}
I(\theta,Y) &\leq I(\theta,X) \label{dpimi}\\
D_Y(p_\theta\|p_{\theta'}) &\leq D_X(p_\theta\|p_{\theta'}).\label{dpidiv}
\end{align}
\end{subequations}
As in Proposition~\ref{prop-dpi}, the first data processing inequality involves a random variable $\theta$, while the second considers $\theta$ as a parameter. The proof is immediate from the chain rules 
$I(\theta; Y) + I(\theta; X|Y)=I(\theta; X,Y) = I(\theta; X)$ and
$D_Y(p_\theta\| p_{\theta'}) + D_{X|Y}(p_\theta\| p_{\theta'})=D_{X,Y}(p_\theta\| p_{\theta'}) = D_{X}(p_\theta\| p_{\theta'})$ where by the Markov chain condition, $I(\theta; Y|X)=0$ and $D_{Y|X}(p_\theta\| p_{\theta'})=0$, respectively.

Comparing the various proofs of the EPI presented above, it is clear that, as already suggested in Zamir's presentation\cite{Zamir98}, estimation theoretic and information theoretic data processing inequalities are strongly related. Also note that in view of~\eqref{kullback}, the lesser known data processing inequality for Fisher information~\eqref{dpifim} is an immediate consequence of the corresponding inequality for divergence~\eqref{dpidiv}. Indeed, dividing both sides of~\eqref{dpidiv} by $\| \theta-\theta'\|^2$ and letting $\theta' \to \theta$ gives~\eqref{dpifim}. It would be interesting to see if the various data processing inequalities (for mutual information, divergence, MMSE, and Fisher information) can be further unified and given a common viewpoint, leading to new insights and applications.

\subsubsection{On the EPI for Discrete Variables}
The above proof of the MII does not require the $(X_i)_i$ to be random vectors with densities. Therefore, it also holds when the random vectors are discrete (finitely or countably) valued. In fact, Verd\'u and Guo\cite{VerduGuo06} used\cite[Lemma~6, App. VII]{GuoShamaiVerdu05} to show that the EPI~\eqref{epi3} also holds in this case, where differential entropies are replaced by entropies. We call attention that this is in fact a immediate consequence of the stronger inequality
$$
H(\sum_i a_i X_i) \geq \max_i H(X_i) %\geq \sum_i a_i^2 H(X_i)
$$
for any independent discrete random vectors $(X_i)_i$ and any real-valued coefficients $(a_i)_i$, which is easily obtained by noting that $H(\sum_i a_i X_i) \geq H(\sum_i a_i X_i| (X_j)_{j\ne i}) = H(X_i)$ for all $i$. Note, however, that the classical EPI in the form 
$\exp \frac{2}{n}H(\sum_i X_i) \geq \sum_i \exp \frac{2}{n}H(X_i)$ does \emph{not} hold in general for discrete random vectors---a simple counterexample is obtained by taking deterministic $X_i$ for all $i$.

There also exist may discrete analogs to the entropy power inequality, either in the form~\eqref{epi1} or~\eqref{epi3}. A first set of results~\cite{WynerZiv73,Wyner73,Witsenhausen74,ShamaiWyner90} were derived for binary random vectors where addition is replaced by modulo-2 addition. The corresponding inequalities are quite different from~\eqref{epi} and apparently unrelated to the contributions of this paper. 

More recent results involve random variables taking integer values. In this case, the role of the Gaussian distribution and it stability property under convolution is played by the Poisson distribution. An analog of the FII~\eqref{fii} was proposed by Kagan~\cite{Kagan01} and a similar, alhtough different, version of discrete Fisher information was used in~\cite{KontoyiannisHarremoesJohnson05} in connection with the convergence of the (usual) sum of independent binary random variables toward the Poisson distribution. A discrete analog to~\eqref{epi1} was proved for binomial distributions~\cite{HarremoesVignat03}, and a discrete analog to~\eqref{epi3} was recently established by Yu and Johnson~\cite{YuJohnson09} for positive random variables having ultra-log-concave distributions. It would be desirable to unify the different approaches for integer-valued random variables to see whether the method of this paper contributes to what is known in this case.

\section{Zamir and Feder's EPI for Linear Transformations}\label{sec-zfepi}

\subsection{Background}

Zamir and Feder\cite{ZamirFeder93, ZamirFeder93a, ZamirFeder93b} generalized the scalar EPI by extending the linear combination $\sum_i a_i X_i$ of random variables to an arbitrary linear transformation $\A X$, where $X$ is the random vector of independent entries $(X_j)_j$ and $\A=(a_{i,j})_{i,j}$ is a rectangular matrix. They showed that the resulting inequality cannot be derived by a straightforward application of the vector EPI of Proposition~\ref{prop-epi}. They also noted that it becomes trivial if $\A$ is row-rank deficient. Therefore, in the following, we assume that $\A$ has full row rank.

Zamir and Feder's generalized EPI (ZF-EPI) has been used to derive results on closeness to normality after linear transformation of a white random vector in the context of minimum entropy deconvolution\cite{ZamirFeder93} and analyze the rate-distortion performance of an entropy-coded dithered quantization scheme~\cite{ZamirFeder95}. It was also used as a guide to extend the Brunn-Minkowski inequality in geometry~\cite{ZamirFeder95a,ZamirFeder98}, which can be applied to the calculation of lattice quantization bit rates under spectral constraints.

The equivalent forms of the ZF-EPI corresponding to those given in Proposition~\ref{prop-epi} are the following.
\begin{proposition}[Equivalent ZF-EPIs]\label{prop-zfepi}
The following inequalities, each stated for any random (column) vector $X$ of independent entries $(X_j)_j$ with densities and real-valued rectangular full row rank matrix $\A$, are equivalent.
\begin{subequations}\label{zfepi}
\begin{align}\label{zfepi1}
N(\A X) &\geq |\A \, \diag (N(X_j))_j \, \A|^{1/r},
\\\label{zfepi2}
h(\A X) &\geq h(\A \widetilde{X}),
\\\label{zfepi3}
h(\A X) &\geq \sum_{i,j} a_{i,j}^2 h(X_j) \qquad (\A\A^{t} = \I),
\end{align}
\end{subequations}
where $r$ is the number of rows in $\A$, and the components of $\widetilde{X}=(\widetilde{X}_j)_j$ are independent Gaussian random variables of entropies $h(\widetilde{X}_j)=h(X_j)$.
\end{proposition}

The proof of Proposition~\ref{prop-zfepi} is a direct extension of that of Proposition~\ref{prop-epi}. That \eqref{zfepi1}, \eqref{zfepi2} are equivalent follows immediately from the equalities $|\A \, \diag (N(X_j))_j \, \A|^{1/r} = |\A \, \diag (N(\widetilde{X}_j))_j \, \A|^{1/r}=|\A \Cov(\widetilde{X})\A^t|^{1/r}=|\Cov(\A \widetilde{X})|^{1/r}=N(\A \widetilde{X})$. The implication \eqref{zfepi3}$\implies$\eqref{zfepi1} is proved in\cite{ZamirFeder93b}, and the equivalence \eqref{zfepi2}$\iff$\eqref{zfepi3} is proved in detail in\cite{GuoShamaiVerdu06a}.

Similarly as for~\eqref{epi3}, inequality~\eqref{zfepi3} can be interpreted as a concavity property of entropy under the variance-preserving\footnote{If the $(X_j)_j$ have equal variances, then so have the components of $\A X$, since $\Cov(X)=\sigma^2 \I$ implies $\Cov(\A X)=\A\Cov(X)\A^t=\sigma^2 \A \A^t = \sigma^2\I$.} transformation
\begin{equation}\label{zftransformation}
X \to \A X \qquad (\A\A^t=\I)
\end{equation}
and is the golden door in the route of proving the ZF-EPI. The conventional techniques presented in Section~\ref{sec-earlier} generalize to the present situation. One has the following Fisher information matrix inequalities analogous to~\eqref{zfepi}:
\begin{subequations}\label{zffiim}
\begin{align}\label{zffiim1}
\J^{-1}(\A X) &\geq \A \J^{-1}(X)\A^t,
\\\label{zffiim2}
\J(\A X) &\leq \J(\A \widehat{X}),
\\\label{zffiim3}
\J(\A X) &\leq \A \J(X) \A^t \qquad (\A\A^{t} = \I),
\end{align}
\end{subequations}
where the components of $\widehat{X}=(\widehat{X}_j)_j$ are independent Gaussian variables with Fisher informations $J(\widehat{X}_j)=J(X_j)$ for all $j$. 
The first inequality~\eqref{zffiim1} was derived by Papathanasiou\cite{Papathanasiou93} and independently by Zamir and Feder\cite{ZamirFeder93,ZamirFeder93b}, who used a generalization of the conditional mean representation of score (see Section~\ref{sec-blachman}); their proof is simplified in\cite{VignatBercher02,VignatBercher03}. Later, Zamir\cite{Zamir98} provided an insightful proof of~\eqref{zffiim} by generalizing Stam's approach (see Section~\ref{sec-stam}) and also determined the case of equality\cite{Zamir97,VignatBercher02}. 
Taking the trace in both sides of~\eqref{zffiim3} gives
\begin{equation}\label{zffii}
J(\A X) \leq \sum_{i,j} a^2_{i,j} J(X_j) \qquad (\A\A^{t} = \I),
\end{equation}
which was used by Zamir and Feder\cite{ZamirFeder93b,Zamir98} to prove the ZF-EPI~\eqref{zfepi3} by integration over the path $\{\sqrt{t}X+\sqrt{1-t}Z\}$ (see Section~\ref{sec-fiiepi}). Finally, Guo, Shamai and Verdu\cite{GuoShamaiVerdu06a} generalized their approach (see Section~\ref{sec-verduguo}) to obtain the inequality 
\begin{equation}\label{zfmmse}
\Var(\A X|\A X+Z)\geq \sum_{i,j} a^2_{i,j} \Var(X_j|X_j+Z_j),
\end{equation}
where $Z$ and the $(Z_j)_j$ are standard Gaussian independent of $X$,
and used it to prove the ZF-EPI~\eqref{zfepi3} by integration over the path $\{\sqrt{t}X+Z\}$ (see Section~\ref{sec-fiiepi}). Again the approaches corresponding to \eqref{zffii} and~\eqref{zfmmse} are equivalent by virtue of the complementary relation~\eqref{fimmsewhite}, as explained in section~\ref{sec-verduguo}.
 
\subsection{A New Proof of the ZF-EPI}

The same ideas as in the proof of Theorem~\ref{th-mii} are easily generalized to prove the ZF-EPI.

\begin{theorem}[Mutual Information Inequality for Linear Transformations]\label{th-zfmii}
For any random vector $X$ with independent entries $(X_j)_j$ having finite variances, any real-valued rectangular matrix $\A$ with $r$ orthonormal rows ($\A\A^t = \I$), and any standard Gaussian random $r$-vector $\Z$ and variable $Z$ independent of $X$, 
\begin{equation}\label{zfmii}
I(\A X + \Z;\Z) \leq \sum_{i,j} a_{i,j}^2 I(X_j+Z;Z).
\end{equation}
Furthermore, this inequality imply the ZF-EPI.
\end{theorem}
\medskip

\begin{proof}
Noting $Z'=\A^t \Z$, a Gaussian random vector with the same dimension as $X$, we can write the following string of inequalities: 
\begin{subequations}
\begin{align}
I(\A X + \Z;\Z) &=I(\A (X + Z');Z') \label{b1}\\
&\leq  I( X + Z' ;Z')  \label{b2}\\
&\leq \sum_j I(X_j + Z'_j ;Z'_j)\label{b3}
\end{align}
\end{subequations}
where~\eqref{b1} holds since $\A\A^t=\I$, \eqref{b2}~follows from the data processing theorem applied to the linear transformation~\eqref{zftransformation}, and \eqref{b3}~follows from Sato's inequality (Lemma~\ref{lem-sato}). Now apply the resulting inequality to $\hat{X}=X+\sqrt{u}\,\hat{Z}$, where $u>0$ and $\hat{Z}$ is a standard Gaussian random vector independent of all other random variables, and replace $\Z$ by $\sqrt{t}\,\Z$, where $t>0$. This gives
$$
I(\A X + \sqrt{u}\, \A\hat{Z} + \sqrt{t}\, \Z; \Z) \leq \sum_j I(X_j + \sqrt{u}\,\hat{Z}_j +\sqrt{t}\,Z'_j;Z'_j).
$$
The Gaussian perturbation $\hat{Z}$ ensures that densities of the $(X_j + \sqrt{u}\,\hat{Z}_j)_j$ are smooth, so that~\eqref{continuity} of Lemma~\ref{lem-continuity} applies to the right-hand side. Noting that $\Cov(Z')=\A^t\A$ and therefore, $\sigma^2_{Z'_j}=\sum_i a^2_{i,j}$ for all $j$, we obtain 
$$
I(\A X + \sqrt{u}\, \A\hat{Z} + \sqrt{t}\, \Z; \Z) \leq \sum_{i,j} a^2_{i,j} I(X_j + \sqrt{u}\,\hat{Z}_j +\sqrt{t}\,Z;Z) + o(t)
$$
where $\A\hat{Z}$ is identically distributed as $\Z$ (since $\A\A^t=\I$), and $Z$ is a standard Gaussian variable, independent of all other random variables. By the stability property of the Gaussian distribution under convolution, $\sqrt{u}\, \A\hat{Z} + \sqrt{t}\, \Z$ is identically distributed as $\sqrt{u+t}\,\Z$, and the $(\sqrt{u}\,\hat{Z}_j +\sqrt{t}\,Z)_j$ are identically distributed as $\sqrt{u+t}\,Z$. Therefore, applying~\eqref{midiff} gives
$$
I(\A X + \sqrt{u+t}\, \Z; \Z) - I(\A X + \sqrt{u}\, \Z; \Z) \leq 
\sum_{i,j} a^2_{i,j} \bigl( I(X_j + \sqrt{u+t}\, Z; Z) - I(X_j + \sqrt{u+t}\, Z; Z)\bigr) + o(t)
$$
which shows that
$$
f(t) =  I(\A X + \sqrt{t}\, \Z; \Z) - \sum_{i,j} a^2_{i,j} I(X_j + \sqrt{t}\, Z; Z)
$$
is nonincreasing in $t>0$. Also, by Lemma~\ref{lem-continuity}, $\lim_{t\to 0} f(t) = f(0)=0$. Therefore, $f(1)\leq f(0)=0$, which proves the required MII~\eqref{zfmii}.

Finally, we show that~\eqref{zfmii} implies the ZF-EPI~Ê\eqref{zfepi3}.
By the identity $I(X+Z;Z)=I(X;X+Z)+h(Z)-h(X)$ for any $X$ independent of $Z$, the MII in the form $f(t)\leq 0$ can be rewritten as
$$
h(\A X ) - \sum_{i,j} a_{i,j}^2 h(X_j) \geq I(\A X; \A X+ \sqrt{t}\,\Z) - \sum_{i,j} a_{i,j}^2 I(X_j;X_j+\sqrt{t}\,Z) + \Delta.
$$
where $\Delta=h(\sqrt{t}\,\Z)-\sum_{i,j} a_{i,j}^2 h(\sqrt{t}\,Z)=r h(\sqrt{t}\,Z) - rh(\sqrt{t}\,Z)=0$.
The other terms in the right-hand side of this inequality are of the form $I(X+\sqrt{t}\,Z;Z)=I(X;\frac{1}{\sqrt{t}}\, X+Z)$, which letting $t\to\infty$ tends to zero by Lemma~\ref{lem-continuity}. This completes the proof.
\end{proof}

Notice that the approach presented here for proving the ZF-EPI is the same as for proving the original EPI, namely, that the difference between both sides of the ZF-EPI~\eqref{zfepi3} is decreased as as independent white Gaussian noise is added:
\begin{equation}\label{zfmii2}
h(\A X ) - \sum_{i,j} a_{i,j}^2 h(X_j) \geq h(\A X') - \sum_{i,j} a_{i,j}^2 h(X'_j),
\end{equation}
where $X'=X+Z$ and $Z$ is white Gaussian independent of $X$.

Zamir and Feder derived their results for random variables $X_j$. However, our approach can be readily extended to random $n$-vectors. For this purpose, consider the random vector $X=(X_j)_j$ whose components $X_j$ are themselves $n$-vectors, and adopt the convention that the components of $Y=\A X$ are $n$-vectors given by the relations $Y_i = \sum_j a_{i,j}X_j$, which amounts to saying that $\A$ is a block matrix with submatrix entries $(a_{i,j}\I)_{i,j}$. The generalization of Theorem~\ref{th-zfmii} is straightforward and we omit the details. The corresponding general ZF-EPI is still given by~\eqref{zfepi}, with the above convention in the notations.

\section{Takano and Johnson's EPI for Dependent Variables}\label{sec-tjepi}

\subsection{Background}

Takano\cite{Takano96} and Johnson\cite{Johnson04} provided conditions under which the EPI, in the form $N(X_1+X_2)\geq N(X_1)+N(X_2)$, would still hold for dependent variables. These conditions are expressed in terms of appropriately perturbed variables
\begin{equation}\label{tjperturbed}
X_{i,t}=X_i+\sqrt{f_i(t)}\,Z_i \qquad (i=1,2)
\end{equation}
where $Z_1,Z_2$ are standard Gaussian, independent of $X=(X_1,X_2)^t$ and of each other, and $f_1(t)$ and $f_2(t)$ are positive functions which tend to infinity as $t\to\infty$. They involve individual scores $S(X_{1,t}),S(X_{2,t})$ and Fisher informations $J(X_{1,t}),J(X_{2,t})$, as well as the entries of the joint score $S(X_t)=\bigl(S_1(X_t), S_2(X_t)\bigr)^t$ and the Fisher information matrix
$$
\J(X_t) = \begin{pmatrix}
J_{1,1}(X_t) & J_{1,2}(X_t)\\
J_{1,2}(X_t) & J_{2,2}(X_t)
\end{pmatrix},
$$
where $X_t=(X_{1,t},X_{2,t})^t$. Takano's condition is\cite{Takano96}
\begin{equation}\label{takano}
2 \frac{\E \bigl(S(X_{1,t})S(X_{2,t})\bigr)}{J(X_{1,t})J(X_{2,t})} \geq
\E \biggl(\Bigl\{ \frac{S_1(X_t)-S(X_{1,t})}{J(X_{1,t})}+\frac{S_2(X_t)-S(X_{2,t})}{J(X_{2,t})}\Bigr\}^2\biggr)
\end{equation}
for all $t>0$. Johnson's improvement is given by the following weaker condition\cite{Johnson04}:
\begin{equation}\label{johnson}
\begin{split}
2 \frac{\E \bigl(S(X_{1,t})S(X_{2,t})\bigr)}{J(X_{1,t})J(X_{2,t})} &\geq\\&\mspace{-72mu}
\E \biggl(\Bigl\{ 
\frac{\bigl(J_{2,2}(X_t)-J_{1,2}(X_t)\bigr)S_1(X_t)+\bigl(J_{1,1}(X_t)-J_{1,2}(X_t)\bigr)S_2(X_t)}{J_{1,1}(X_t)J_{2,2}(X_t)-J^2_{1,2}(X_t)}
-\frac{S(X_{1,t})}{J(X_{1,t})}-\frac{S(X_{2,t})}{J(X_{2,t})}\Bigr\}^2\biggr)
\end{split}
\end{equation}
for all $t>0$. These conditions were found by generalizing the conventional approach presented in Section~\ref{sec-earlier}, in particular Blachman's representation of the score (Section~\ref{sec-blachman}). They are simplified below. The EPI for dependent variables finds its application in entropy-based blind source separation of dependent components (see e.g.,~\cite{CaiafaKuruogluProto06}).

\subsection{A Generalized EPI for Dependent Random Vectors}

In this section, we extend Theorem~\ref{th-mii} to provide a simple condition on dependent random $n$-vectors $(X_i)_i$ under which not only the original EPI $N(\sum_i X_i)\geq \sum_i N(X_i)$ holds, but also the EPI~\eqref{epi} for  \emph{any} choice of coefficients $(a_i)_i$. Such stronger form should be more relevant in applications such as blind separation of dependent components, for it ensures that negentropy $-h$ still satisfies the requirements~\eqref{contrast} for a contrast objective function, for any type of linear mixture.
Define
\begin{equation}\label{perturbed}
X_{i,t}=X_i+\sqrt{t}\,Z_i
\end{equation}
corresponding to~\eqref{tjperturbed} with $f_i(t)=t$ for all $i$. Our condition will be expressed in terms of \emph{symmetric mutual information} $I\{(X_{i,t})_i\}\geq 0$ defined by~\eqref{symmetricmi}, which serves as a measure of dependence between the components of a random vector. 

\begin{theorem}\label{th-tjepi}
Let $X=(X_i)_i$ be any finite set of (dependent) random $n$-vectors, let $X_t=(X_{i,t})_i$ be defined by~\eqref{perturbed}, and let $Z$ be a white Gaussian random $n$-vector independent of all other random vectors. If, for any $t>0$ and any real-valued coefficients $(a_i)_i$, adding a small perturbation $a_i Z$ to the $X_{i,t}$ makes them ``more dependent'' in the sense that
\begin{equation}\label{condition}
I\{(X_{i,t}+a_i Z)_i\} \geq I\{(X_{i,t})_i\} + o(\sigma^2_Z)
\end{equation}
then the MII~\eqref{mii} and the EPI~\eqref{epi} hold for these random vectors $(X_i)_i$.
\end{theorem}

\begin{proof}
The only place where the independence of the $(X_i)_i$ is used in the proof of Theorem~\ref{th-mii} is Sato's inequality~\eqref{a3}, which is used to the first order of $\sigma^2_Z$ and applied to random vectors of the form~\eqref{perturbed} for all $t>0$. Therefore it is sufficient that
$$
I( (X_{i,t} + a_i Z)_i ;Z)\leq \sum_i I(X_{i,t} + a_i Z ;Z) + o(\sigma^2_Z)
$$
holds for all $t>0$ and any choice of $(a_i)_i$ to prove the MII and hence the EPI.
Now from the proof of Lemma~\ref{lem-sato}, the difference between both sides of this inequality is
$$
I( (X_i + a_i Z)_i ;Z)- \sum_i I(X_i + a_i Z ;Z) =I\{(X_i)_i\}- I\{(X_i+a_i Z)_i\}
$$
The result follows at once.
\end{proof}
Note that it is possible to check~\eqref{condition} for a \emph{fixed} choice of the coefficients $(a_i)_i$ to ensure that the EPI~\eqref{epi3} holds for these coefficients.
Of course,~\eqref{condition} is obviously always satisfied for independent random vectors $(X_i)_i$. In order to relate condition~\eqref{condition} to Takano and Johnson's~\eqref{takano}, \eqref{johnson}, we rewrite the former in terms of Fisher information as follows.

\begin{corollary}
For random variables $(X_i)_i$ ($n=1$), condition~\eqref{condition} is equivalent to
\begin{equation}\label{conditionfisher}
\diag\bigl(J(X_{i,t})\bigr)_i\geq \J(X_t)
\end{equation}
for all $t>0$, where $X_t=(X_{i,t})_i$. Therefore, if this condition is satisfied then the MII~\eqref{mii} and the EPI~\eqref{epi} hold.
\end{corollary}

\begin{proof}
Let $Z$ be a standard Gaussian random variable independent of $X_t$, and define $a=(a_i)_i$ and $Y_t=X_t+\sqrt{\eps}\, a Z$, where $aZ=(a_i Z)_i$. The perturbations~\eqref{perturbed} ensure that the density of $X_t$ is smooth, so that the function $I(\eps)=I\{(Y_{i,t})_i\}$ is differentiable for all $\eps\geq 0$. Now condition~\eqref{condition} is equivalent to the inequality $I(\eps)\geq I(0)+o(\eps)$, that is, $I'(0)\geq 0$. By definition~\eqref{symmetricmi}, 
$$
I\{(Y_{i,t})_i\} = \sum_i h(Y_{i,t}) - h(Y_t),
$$
so the inequality $I'(0)\geq 0$ can be rewritten as
$$
\frac{d}{d\eps} \sum_i h(X_{i,t}+\sqrt{\eps}\, a_i Z) - h(X_t+\sqrt{\eps}\, aZ)\Bigr|_{\eps=0} \geq 0.
$$
By de Bruijn's identity~\eqref{debruijn}, this is equivalent to
$$
\sum_i a^2_i J(X_{i,t}) \geq  \tr \bigl( \J(X_t) \Cov(aZ) \bigr)
$$
where $\Cov(aZ)=a\Var(Z) a^t = aa^t$, that is,
\begin{equation}\label{conditionfisher2}
a^t \cdot \diag\bigl(J(X_{i;t})\bigr)_i \cdot a \geq a^t\cdot  \J(X_t) \cdot a
\end{equation}
for any vector $a$ and $t>0$. This shows that \eqref{condition} is equivalent to the matrix inequality~\eqref{conditionfisher} as required.
\end{proof}
We now recover Takano and Johnson's conditions~\eqref{takano},~\eqref{johnson} from~\eqref{conditionfisher}.

\begin{lemma}
In the case of two random variables $X_1,X_2$, conditions~\eqref{takano} and \eqref{johnson} are equivalent to
\begin{align}
{\lambda}^t \cdot \diag\bigl(J(X_{i;t})\bigr)_i \cdot \lambda \geq {\lambda}^t\cdot  \J(X_t) \cdot \lambda \label{c1}\\
{\lambda}^t \cdot \diag\bigl(J(X_{i;t})\bigr)_i \cdot \lambda \geq {\mu}^t\cdot  \J(X_t) \cdot \mu, \label{c2}
\end{align}
respectively, where $\lambda$ and $\mu$ minimize the quadratic forms ${a}^t \cdot \diag\bigl(J(X_{i;t})\bigr)_i \cdot a$ and ${a}^t\cdot  \J(X_t) \cdot a$, respectively, over all vectors $a$ of the form $a=(\alpha, 1-\alpha)^t$, $0\leq \alpha\leq 1$.
\end{lemma}

\begin{proof}
Given a  positive definite symmetric matrix $\J$, the general solution $a^*$ to
$$
\min_a \; a^t\cdot \J \cdot a \qquad (a_i\geq 0, \sum_i a_i=1)
$$
is easily found by the Lagrangian multiplier method. One finds $a^*_i=\sum_{j} J^{-1}_{i,j}/\sum_{i,j} J^{-1}_{i,j}$ for all $i$ and 
${a^*}^t\cdot \J \cdot a^* = \bigl(\sum_{i,j} J^{-1}_{i,j}\bigr)^{-1}$,
where $J^{-1}_{i,j}$ are the entries of the inverse matrix $\J^{-1}$.
Particularizing this gives $\lambda_1\propto J^{-1}(X_{1,t})$, $\lambda_2\propto  J^{-1}(X_{2,t})$ and $\mu_1\propto J_{2,2}(X_t)-J_{1,2}(X_t)$, $\mu_2\propto J_{1,1}(X_t)-J_{1,2}(X_t)$ up to appropriate proportionality factors, and~\eqref{c1},~\eqref{c2} are rewritten as
\begin{align}
J^{-1}(X_{1,t})+J^{-1}(X_{2,t}) &\geq \frac{J_{1,1}(X_t)}{J^2(X_{1,t})}+\frac{J_{2,2}(X_t)}{J^2(X_{2,t})}+2 \frac{J_{1,2}(X_t)}{J(X_{1,t})J(X_{2,t})}  \label{c3}
\\
\bigl(J^{-1}(X_{1,t})+J^{-1}(X_{2,t})\bigr)^{-1} &\geq \frac{J_{1,1}(X_t)J_{2,2}(X_t)-J^2_{1,2}(X_t)}{J_{1,1}(X_t)+J_{2,2}(X_t)-2J_{1,2}(X_t)}   \label{c4}
\end{align}
Meanwhile, expanding the right-hand sides in~\eqref{takano}, \eqref{johnson} using Stein's identity\cite{Johnson04} gives 
$$
2(\nu_1+\nu_2) - J^{-1}(X_{1,t}) - J^{-1}(X_{2,t})
\geq 
\nu^2_1 J_{1,1}(X_t) +\nu^2_2 J_{2,2}(X_t)+2\nu_1\nu_2 J_{1,2}(X_t),
$$
where $(\nu_1,\nu_2)=\bigl(J^{-1}(X_{1,t}),J^{-1}(X_{2,t})\bigr)$ for Takano's condition and $(\nu_1,\nu_2)=\bigl(J_{2,2}(X_t)-J_{1,2}(X_t),J_{1,1}(X_t)-J_{1,2}(X_t)\bigr) / \bigl(J_{1,1}(X_t)J_{2,2}(X_t)-J^2_{1,2}(X_t)\bigr)$ for Johnson's condition. Replacing yields~\eqref{c3} and~\eqref{c4}, respectively. This proves the lemma.
\end{proof}

\begin{corollary}
In the case of two random variables $X_1,X_2$, condition~\eqref{conditionfisher} implies both Takano and Johnson's conditions~\eqref{takano}, \eqref{johnson}. 
\end{corollary}

\begin{proof}
Condition~\eqref{conditionfisher} implies~\eqref{conditionfisher2} for any $a$ of the form $a=(\alpha, 1-\alpha)^t$, $0\leq \alpha\leq 1$. Setting $a=\lambda$ yields Takano's condition~\eqref{c1}. Replacing the right-hand side of the resulting inequality by the minimum over $a$ (achieved by $a=\mu$) gives Johnson's condition~\eqref{c2}.
\end{proof}

Thus, our condition~\eqref{conditionfisher} is stronger than Takano's or Johnson's. This is not surprising since it yields a stronger form of the EPI~\eqref{epi}, valid for any choice of coefficients~$(a_i)_i$.

\section{Liu and Viswanath's covariance-constrained EPI}\label{sec-lvepi}

As already mentioned in the introduction, all known applications of the EPI to source and channel coding problems\cite{Bergmans74,WeingartenSteinbergShamai06, MohseniCioffi06,Leung-Yan-CheongHellman78,TekinYener06,TekinYener07,Costa85a ,Ozarow80, Zamir99,Oohama97,Oohama98,Oohama05,Oohama06} involve an inequality of the form 
$N(X+Z)\geq N(X)+N(Z)$, where $Z$ is Gaussian independent of $X$. In this and the next section, we study generalizations of this inequality. We begin with Liu and Viswanath's generalized EPI for constrained covariance matrices.

\subsection{Background}

Recently, Liu and Viswanath\cite{LiuViswanath06,LiuViswanath07} have suggested that the EPI's main contribution to multiterminal coding problems is for solving optimization problems of the form
\begin{equation}\label{lvproblem}
\max_{p(x)} h(X)-\mu h(X+Z) \qquad (\mu\geq 1)
\end{equation}
where $Z$ is Gaussian and the maximization is over all random $n$-vectors $X$ independent of $Z$. The solution is easily determined from the EPI in the form~\eqref{epi3} applied to the random vectors $X_1=\mu^{1/2}X$ and $X_2=(1-\mu^{-1})^{-1/2}Z$:
$$
h(X+Z)\geq \mu^{-1} \bigl(h(X)+\frac{n}{2}\log\mu\bigr)+(1-\mu^{-1}) h\bigl((1-\mu^{-1})^{-1/2}Z\bigr).
$$
Since equality holds iff $X_1$ and $X_2$ are Gaussian with identical covariances, it follows that the optimal solution $X$ to~\eqref{lvproblem} is Gaussian with covariance matrix $\Cov(X)=(\mu-1)^{-1}\Cov(Z)$. 

Clearly, the existence of a Gaussian solution to~\eqref{lvproblem} is equivalent to the EPI for two independent random vectors $X$ and $Z$. Liu and Viswanath\cite{LiuViswanath06,LiuViswanath07} have found an implicit generalization of the EPI by showing that~\eqref{lvproblem} still admits a Gaussian solution under the covariance constraint $\Cov(X)\leq \C$, where $\C$ is any positive definite matrix. 
The gave a ``direct proof,'' motivated by the vector Gaussian broadcast channel problem, using the classical EPI, the saddlepoint property of mutual information~\eqref{saddle} and the ``enhancement'' technique for Gaussian random vectors introduced by Weingarten, Steinberg and Shamai\cite{WeingartenSteinbergShamai06}. They also gave a ``perturbation proof'' using a generalization of the conventional techniques presented in Section~\ref{sec-earlier}, namely, an integration over a path of the form $\{\sqrt{1-t}\,X+\sqrt{t}\,Z\}$ of a generalized FII~\eqref{fii3} with matrix coefficients, using de Bruijn's identity and the Cram\'er-Rao inequality\footnote{As explained in Section~\ref{sec-saddle}, the Cram\'er-Rao inequality~\eqref{crb} is equivalent to the saddlepoint property~\eqref{saddle} used in their ``direct proof''.}. This and similar results for various optimization problems involving several Gaussian random vectors find applications in vector Gaussian broadcast channels and distributed vector Gaussian source coding\cite{LiuViswanath07}.

\subsection{An Explicit Covariance-Constrained MII}

We first give explicit forms of covariance-constrained MII and EPI, which will be used to solve Liu and Viswanath's optimization problem. Again, the same ideas as in the proof of Theorem~\ref{th-mii} are easily generalized to prove the following covariance-constrained MII and EPI, using only basic properties of mutual information.

\begin{theorem}\label{th-lvmii}
Let $X_1,X_2$ be independent random $n$-vectors with positive definite covariance matrices, and let $Z_1,Z_2$ be Gaussian random $n$-vectors independent of $X_1,X_2$ and of each other, with covariances proportional to those of $X_1$ and $X_2$, respectively: $\Cov(Z_1)=\alpha \Cov(X_1)$, $\Cov(Z_2)=\alpha \Cov(X_2)$, where $\alpha>0$. Assume that $X_2$ is Gaussian and $X_1,X_2$ are subject to the covariance constraint
\begin{equation}\label{covconstraint}
\Cov(X_1)\leq \Cov(X_2).
\end{equation}
Then for any real-valued coefficients $a_1,a_2$ normalized such that $a_1^2+a_2^2=1$,
\begin{equation}\label{lvmii}
I(a_1X_1+a_2X_2+Z;Z)\leq a_1^2 I(X_1+Z_1;Z_1)+a_2^2I(X_2+Z_2;Z_2)
\end{equation}
where we have noted $Z=a_1Z_1+a_2Z_2$. Furthermore, this inequality implies the following generalized EPI:
\begin{equation}\label{lvepi}
h(a_1X_1+a_2X_2)\geq a^2_1h(X_1)+a_2^2h(X_2) + \Delta
\end{equation}
where 
\begin{equation}\label{lvdelta}
\Delta=h(Z)- a^2_1h(Z_1)-a_2^2h(Z_2)\geq 0.
\end{equation}
\end{theorem}

Note that for the particular case $\Cov(X_1)=\Cov(X_1)$, we have $\Cov(Z_1)=\Cov(Z_2)$, the random vectors $Z_1,Z_2$ and $Z$ are identically distributed, $\Delta=0$ and Theorem~\ref{th-lvmii} reduces to Theorem~\ref{th-mii} for two random vectors.

\begin{proof}[Proof of Theorem~\ref{th-lvmii}]
First define
$$
Z'_i=\Cov(Z_i)\Cov(Z)^{-1}Z \qquad (i=1,2)
$$
with covariance matrices $\Cov(Z'_i)=\Cov(Z_i)\Cov(Z)^{-1}\Cov(Z_i)$. From~\eqref{covconstraint}, one successively has $\Cov(Z_1)\leq \Cov(Z_2)$, $\Cov(Z_1)\leq a_1^2\Cov(Z_1)+a_2^2\Cov(Z_2)= \Cov(Z)$, $\Cov(Z)^{-1}\leq \Cov^{-1}(Z_1)$, and upon left and right multiplication by $\Cov(Z_1)$, $\Cov(Z'_1)=\Cov(Z_1)\Cov(Z)^{-1}\Cov(Z_1)\leq \Cov(Z_1)$. Similarly, $\Cov(Z'_2)\geq \Cov(Z_2)$. Therefore, we can write
\begin{equation}\label{tilde}
\begin{aligned}
Z_1&=Z'_1+\widetilde{Z}_1 \\
Z'_2&=Z_2+\widetilde{Z}_2 \\
\end{aligned}
\end{equation}
where $\widetilde{Z}_1$ and $\widetilde{Z}_1$ are Gaussian and independent of $Z'_1$ and $Z_2$, respectively. We can now write the following string of inequalities:
\begin{subequations}
\begin{align}
I(a_1X_1+a_2X_2+Z;Z) &= I(a_1(X_1+a_1Z'_1)+a_2(X_2+a_2Z'_2);Z)\label{v1}\\
&\leq I(X_1+a_1 Z'_1,X_2+a_2 Z'_2;Z)\label{v2}\\
&\leq I(X_1+a_1 Z'_1;Z'_1)+I(X_2+a_2 Z'_2;Z'_2)\label{v3}\\
&= I(X_1+a_1Z_1;Z_1)+I(X_2+a_2Z_2;Z_2)\nonumber\\
 &\quad - I(X_1+a_1(Z'_1+\widetilde{Z}_1); \widetilde{Z}_1) + I(X_2+a_2(Z_2+\widetilde{Z}_2); \widetilde{Z}_2) \label{v4}
\end{align}
\end{subequations}
where~\eqref{v1} holds since $a_1^2Z'_1+a_2^2Z'_2=\Cov(Z)\Cov(Z)^{-1}Z=Z$, \eqref{v2}~follows from the data processing theorem applied to the linear transformation~\eqref{transformation}, \eqref{v3}~follows from Sato's inequality (Lemma~\ref{lem-sato}), and \eqref{v4}~follows by applying the identity~\eqref{midiff} to the random vectors defined by~\eqref{tilde}.
By Proposition~\ref{prop-saddle}, $I(X_1+a_1(Z'_1+\widetilde{Z}_1); \widetilde{Z}_1)\geq I(X^*_1+a_1(Z'_1+\widetilde{Z}_1); \widetilde{Z}_1)$, where $X^*_1$ is Gaussian with covariance matrix $\Cov(X^*_1)=\Cov(X_1)$.

We now use the assumption that $\Cov(Z_1)=\alpha \Cov(X_1)$, $\Cov(Z_2)=\alpha \Cov(X_2)$ and let $\alpha\to 0$ in the well-known expressions for mutual informations of Gaussian random vectors:
\begin{subequations}
\begin{align*}
I(X^*_1+a_1(Z'_1+\widetilde{Z}_1); \widetilde{Z}_1) &= 
\frac{1}{2} \log | \I + \alpha a_1^2 \Cov(\widetilde{Z}_1)\Cov(Z_1)^{-1} | + o(\alpha)
\\
I(X_2+a_2(Z_2+\widetilde{Z}_2); \widetilde{Z}_2) &=
\frac{1}{2} \log | \I + \alpha a_2^2 \Cov(\widetilde{Z}_2)\Cov(Z_2)^{-1} | + o(\alpha)
\end{align*}
\end{subequations}
where $\Cov(\widetilde{Z}_1)=\Cov(Z_1)-\Cov(Z'_1)=\Cov(Z_1)-\Cov(Z_1)\Cov(Z)^{-1}\Cov(Z_1)$ and $\Cov(\widetilde{Z}_2)=\Cov(Z'_2)-\Cov(Z_2)=\Cov(Z_2)\Cov(Z)^{-1}\Cov(Z_2)-\Cov(Z_2)$.
Since $\Cov(Z)=a_1^2\Cov(Z_1)+a_2^2\Cov(Z_2)$ and $a_1^2+a_2^2=1$, we have $a_1^2 \Cov(\widetilde{Z}_1)\Cov(Z_1)^{-1}=a_1^2 (\I - \Cov(Z_1)\Cov(Z)^{-1}) = a_2^2 (\Cov(Z_2)\Cov(Z)^{-1}-\I)=a_2^2 \Cov(\widetilde{Z}_2)\Cov(Z_2)^{-1}$, and therefore,
\begin{align*}
I(X_1+a_1(Z'_1+\widetilde{Z}_1); \widetilde{Z}_1)&\geq I(X^*_1+a_1(Z'_1+\widetilde{Z}_1); \widetilde{Z}_1)\\
&= I(X_2+a_2(Z_2+\widetilde{Z}_2); \widetilde{Z}_2) + o(\alpha).
\end{align*}
It follows from~\eqref{v4} that
\begin{equation}\label{v5}
I(a_1X_1+a_2X_2+Z;Z) \leq I(X_1+a_1Z_1;Z_1)+I(X_2+a_2Z_2;Z_2) + o(\alpha)
\end{equation}
The rest of the proof is entirely similar to that of Theorem~\ref{th-mii}. Here is a sketch. Write~\eqref{v5} for $\widehat{X}_1=X_1+\sqrt{t}\,\widehat{Z}_1$ and $\widehat{X}_2=X_2+\sqrt{t}\,\widehat{Z}_2$, where $\widehat{Z}_i$ is identically distributed as $Z_i$ and independent of all other random vectors, for $i=1,2$. Applying Lemma~\ref{lem-continuity} to the right-hand side of the resulting inequality, this gives
$$
I(a_1X_1+a_2X_2+\sqrt{t}\widehat{Z}+\sqrt{\eps}Z;Z) \leq a^2_1 I(X_1+\sqrt{t}\widehat{Z}_1+\sqrt{\eps}\,Z_1;Z_1)+a^2_2 I(X_2+\sqrt{t}\,\widehat{Z}_2+\sqrt{\eps}\,Z_2;Z_2) + o(\eps)
$$
where $\widehat{Z}$ is identically distributed as $Z$. By virtue of~\eqref{midiff}, this can be written in the form $f(t+\eps)\leq f(t)+o(\eps)$, where
$$
f(t)=I(a_1X_1+a_2X_2+\sqrt{t}\,Z;Z) - a^2_1 I(X_1+\sqrt{t}\,Z_1;Z_1)-a^2_2 I(X_2+\sqrt{t}\, Z_2;Z_2).
$$
Therefore, $f(t)$ is nonincreasing, and $f(1)\leq f(0)$, which is the required MII~\eqref{lvmii}. This in turn can be rewritten in the form
$$ 
h(a_1X_1+a_2X_2)- a^2_1h(X_1)-a_2^2h(X_2) \geq 
\Delta + \Delta'
$$
where $\Delta$ is defined by~\eqref{lvdelta} and $\Delta'=I(a_1X_1+a_2X_2; a_1X_1+a_2X_2+Z)- a^2_1I(X_1;X_1+Z_1)-a_2^2I(X_2;X_2+Z_2)$ tends to zero as $\alpha\to\infty$ by Lemma~\ref{lem-continuity}. This proves~\eqref{lvepi} and the theorem.
\end{proof}

It is now easy to recover Liu and Viswanath's formulation.

\begin{corollary}[Liu and Viswanath\cite{LiuViswanath06,LiuViswanath07}]
The maximization problem~\eqref{lvproblem}, subject to the covariance constraint $\Cov(X)\leq \C$, admits a Gaussian optimal solution $X^*$.
\end{corollary}

\begin{proof}
Let $X^*$ be the optimal solution to the maximization problem obtained by restricting the solution space within Gaussian distributions. Thus $\Cov(X^*)>0$ maximizes
$$
\frac{1}{2}\log \bigl((2\pi e)^n |\Cov(X)|\bigr)-\frac{\mu}{2}\log \bigl((2\pi e)^n |\Cov(X)+\Cov(Z)|\bigr)
$$
over all covariance matrices $\Cov(X)\leq \C$. As stated in\cite{LiuViswanath07} and shown in\cite{WeingartenSteinbergShamai06}, $\Cov(X^*)$ must satisfy the Karush-Kuhn-Tucker condition
$$
\frac{1}{2} \Cov(X^*)^{-1} = \frac{\mu}{2} \bigl( \Cov(X^*)+\Cov(Z)\bigr)^{-1} + \mathbf{M},
$$
where $\mathbf{M}\geq 0$ is a Lagrange multiplier corresponding to the contsraint $\Cov(X)\leq \C$. It follows that $\Cov(X^*)^{-1}\geq \mu \bigl( \Cov(X^*)+\Cov(Z)\bigr)^{-1}$, that is, $\mu\Cov(X^*)\leq \Cov(X^*)+\Cov(Z)$, or $\Cov\bigl(\mu^{1/2}X^*\bigr)\leq \Cov\bigl((1-\mu^{-1})^{-1/2}Z\bigr)$.

Now let $X$ be any random vector independent of $Z$, such that $\Cov(X)=\Cov(X^*)$. Define $a_1=\mu^{-1/2}$, $a_2=(1-\mu^{-1})^{1/2}$, $X_1=\mu^{1/2}X$, $X_2=(1-\mu^{-1})^{-1/2}Z$ and $Z_1=\mu^{1/2}X^*$, and let $Z_2$ be a Gaussian random vector identically distributed as $X_2$ and independent of $Z_1$.
Since $a_1^2+a_2^2=1$ and $\Cov(X_1)=\Cov(Z_1)\leq \Cov(X_2)=\Cov(Z_2)$, we may apply Theorem~Ê\ref{th-lvmii}. By~\eqref{lvepi}, we obtain
$$
h(a_1X_1+a_2X_2)-a_1^2 h(X_1) \geq h(a_1Z_1+a_2Z_2)-a_1^2 h(Z_1),
$$
that is, replacing and rearranging,
$$
h(X)-\mu h(X+Z) \leq h(X^*)-\mu h(X^*+Z).
$$
Therefore, the Gaussian random vector $X^*$ is an optimal solution to~\eqref{lvproblem} subject to the constraint $\Cov(X)\leq \C$. This completes the proof.
\end{proof}

\section{Costa's EPI: Concavity of Entropy Power}\label{sec-cepi}

\subsection{Background}

Costa\cite{Costa85} has strengthened the EPI for two random vectors $X,Z$ in the case where $Z$ is white Gaussian. While it can be easily shown\cite{Costa85,DemboCoverThomas91} that Shannon's EPI for $X,Z$ is equivalent to the inequality 
$$
\frac{d}{dt} N(X+\sqrt{t}\, Z)\geq 1,  %\Big|_{t=0} 
$$
Costa's EPI is the convexity inequality which expresses that the entropy power is a concave function of the power of the added Gaussian noise:
\begin{equation}\label{costa}
\frac{d^2}{dt^2} N(X+\sqrt{t}\, Z) \leq 0 % \Big|_{t=0}
\end{equation}
Alternatively, the concavity of the entropy power is equivalent to saying that the slope
$\delta(t) = \bigl(N(X+\sqrt{t}\, Z)-N(X)\bigr)/t$ drawn from the origin is nonincreasing, while the corresponding Shannon's EPI is weaker, being simply equivalent to the inequality $\delta(1)\geq \delta(\infty) = N(Z)$.

The original proof of Costa through an explicit calculation of the second derivative in~\eqref{costa} is quite involved\cite{Costa85}. His calculations are simplified in\cite{Villani00}. Dembo gave an elegant proof using the FII over the path $\{X+\sqrt{t}Z\}$\cite{Dembo89,DemboCoverThomas91}. Recently, Guo, Shamai, and Verd\'u provided a clever proof using the MMSE over the path $\{\sqrt{t}X+Z\}$\cite{GuoShamaiVerdu06a}.

Costa's EPI has been used to determine the capacity region of the Gaussian interference channel~\cite{Costa85a}. It was also used as a continuity argument about entropy that was required for the analysis of the capacity of flat-fading channels in~\cite{LapidothMoser03}.

\subsection{A New Proof of the Concavity of the Entropy Power}

In his original presentation\cite{Costa85}, Costa proposed the concavity property $N(X+\sqrt{t}\,Z)\geq (1-t) N(X)+ t N(X+Z)$ in the segment $(0,1)$ for white Gaussian $Z$, in which case he showed its equivalence to~\eqref{costa}. He also established this inequality in the dual case where $X$ is Gaussian and $Z$ is arbitrary. In the latter case, however, this inequality is not sufficient to prove that $N(X+\sqrt{t}\,Z)$ is a concave function of $t>0$.
In this section, we prove a slight generalization of Costa's EPI, showing concavity in both cases, for an arbitrary (not necessarily white) Gaussian random vector. Again the proposed proof relies only on the basic properties of mutual information.

\begin{theorem}[Concavity of Entropy Power]
Let $X$ and $Z$ be any two independent random $n$-vectors. If either $X$ or $Z$ is Gaussian, then \mbox{$N(X+\sqrt{t}\, Z)$} is a concave function of $t$.
\end{theorem}

\begin{proof}
To simplify the notation, let $Z_t=\sqrt{t}\, Z$. First, it is sufficient to prove concavity in the case where $Z$ is Gaussian, because, as it is easily checked, the functions $n(t)=N(X+Z_t)$ and $t\cdot n(1/t)=N(X_t+Z)$ are always simultaneously concave.
Next define
$$
f_X(t) = \frac{N(X+Z_t)}{N(X)} =  \exp {\frac{2}{n} I(X+Z_t;Z)}.
$$
Our aim is to prove~\eqref{costa}, that is, $f''_X(t)\leq 0$. Consider the MII~\eqref{mii} in the form
$$
I(X_\lambda + Y_{1-\lambda} + Z_t; Z) \leq \lambda I(X+Z_t;Z) + (1-\lambda) I(Y+Z_t;Z)
$$
where $Y$ is independent of $X,Z$ and $0\leq \lambda\leq 1$. Replacing  $X_\lambda, Y_\lambda$ by $X,Y$ gives the alternative form
$$
I(X + Y + Z_t; Z) \leq \lambda I(X+Z_{\lambda t};Z) + (1-\lambda) I(Y+Z_{(1-\lambda)t};Z)
$$
Choose $Z'$ and $Z''$ such that $Z,Z'$ and $Z''$ are i.i.d. and independent of $X$, and replace $X$ by $X+Z'_u$ and $Y$ by $Z''_v$:
\begin{equation}\label{d1}
I(X + Z'_u + Z''_v + Z_t; Z) \leq \lambda I(X+Z'_u+Z_{\lambda t};Z) + (1-\lambda) I(Z''_v+Z_{(1-\lambda)t};Z).
\end{equation}
We now turn this into a ``mutual information power inequality'' similarly as the EPI~\eqref{epi1} is derived from~\eqref{epi3} in the proof of Proposition~\ref{prop-epi}. Define $M_t(X)$ as the power of a Gaussian random vector $\widetilde{X}$ having covariances proportional to those of $Z$ and identical mutual information $I(X+Z_t;Z)$. By Shannon's capacity formula, $I(\widetilde{X}+Z_t;Z)=\frac{n}{2}\log(1+t\sigma_Z^2/\sigma^2_{\widetilde{X}})$, and therefore
$$
M_t(X) = \frac{t\sigma^2_Z }{ \exp {\frac{2}{n} I(X+Z_t;Z)} -1} = \frac{t\sigma^2_Z}{f_X(t)-1}.
$$
Choose $\lambda\in[0,1]$ such that $I(X+Z'_u+Z_{\lambda t};Z) = I(Z''_v+Z_{(1-\lambda)t};Z)$ in~\eqref{d1}. This is always possible, because the difference has opposite signs for $\lambda=0$ and $\lambda=1$. By applying the function $(\exp(\frac{2}{n}\cdot)-1)^{-1}$ to both sides of~\eqref{d1}, we find the inequality
$$
M_t(X+Z'_u+Z''_v) \geq M_{\lambda t}(X+Z'_u) + v \sigma^2_Z
$$
We now let $t\to 0$ (so that $\lambda t\to 0$). Since $f_{X+Z'_u}(t)=f_X(t+u)/f_X(u)$, and similarly, $f_{X+Z'_u+Z''_v}(t)=f_X(t+u+v)/f_X(u+v)$, we obtain
$$
\frac{f_X(u+v)}{f'_X(u+v)} \geq \frac{f_X(u)}{f'_X(u)} + v
$$
Dividing by $v$ and letting $v\to 0$ gives 
$$
\frac{d}{du} \frac{f_X(u)}{f'_X(u)}\geq 1,
$$
that is, carrying out the derivation, $f_X(u)f_X''(u)/f'_X(u)^2 \leq 0$ or $f''_X(u)\leq 0$ as required.
\end{proof}

It would be interesting to know whether this proof can be adapted to the recent  generalization of Costa's EPI~\cite{PayaroPalomar08,PayaroPalomar09,LiuLiuPoorShamai10} in which $t$ is replaced by an arbitrary positive semi-definite matrix.

\section{Open Questions}\label{sec-open}

\subsection{EPI, FII and MII for Subsets of Independent Variables}

Recently, Artstein, Ball, Barthe, and Naor\cite{ABBN04} proved a new entropy power inequality involving entropy powers of sums of all independent variables excluding one, which solved a long-standing conjecture about the monotonicity of entropy. This was generalized to arbitrary collections of subsets of independent variables (or vectors) by Madiman and Barron\cite{MadimanBarron06,MadimanBarron07}. The generalization of the classical formulation of the EPI takes the form
\begin{equation}\label{abbn1}
N(\sum_i X_i) \geq \frac{1}{k} \sum_S N(\sum_{i\in S} X_i)
\end{equation}
where the sum in the right-hand side is over arbitrary subsets $S$ of indexes, and $k$ is the maximum number of subsets in which one index appears.
Note that we may always assume that subsets $S$ are ``balanced''\cite{MadimanBarron07}, i.e., each index $i$ appears in the right-hand side of~\eqref{abbn1} exactly $k$ times. This is because it is always possible to add singletons to a given collection of subsets until the balancing condition is met; since the EPI~\eqref{abbn1} would hold for the augmented collection, it a fortiori holds for the initial collection as well.

For balanced subsets, the inequalities generalizing~\eqref{epi3},~\eqref{fii3},~\eqref{mmsei} and~\eqref{mii} are the following.
\begin{proposition}
Let $(X_i)_i$ be finitely many random $n$-vectors, let $Z$ be any Gaussian random $n$-vector independent of $(X_i)_i$, and let $(a_i)_i$ be any real-valued coefficients normalized such that $\sum_i a^2_i=1$. Then, for any collection $\{S\}$ of balanced subsets of indexes, 
\begin{subequations}\label{abbn}
\begin{align}\label{abbnfii}
J(\sum_i a_i X_i) &\leq \sum_S a^2_S\; J(X_S),
\\\label{abbnmmsei}
\Var(\sum_i a_i X_i | \sum_i a_i X_i +Z)
&\geq \sum_S a^2_S  \Var( X_S | X_S +Z),
\\\label{abbn3}
h(\sum_i a_i X_i) &\geq \sum_S a^2_S\; h(X_S),
\\\label{abbnmii}
I(\sum_i a_i X_i+Z;Z) &\leq \sum_S a^2_S\; I(X_S+Z;Z),
\end{align}
\end{subequations}
where $a^2_S=\frac{1}{k}\sum_{i\in S} a^2_i$ (so that $\sum_S a^2_S=1$) and $X_S$ is given by the covariance preserving transformation
$$
X_S = \frac{\sum_{i\in S} a_i X_i}{\sqrt{\sum_{i\in S} a^2_i}}.
$$
\end{proposition}
\medskip
Available proofs of~\eqref{abbnfii}--\eqref{abbn3} are generalizations of the conventional techniques presented in Section~\ref{sec-earlier}, where an additional tool (``variance drop lemma'') is needed to prove either~\eqref{abbnfii} or~\eqref{abbnmmsei}; see~\cite[Lemma 5]{ABBN04}, \cite[Lemma3]{TulinoVerdu06}, or \cite[Lemma 2]{MadimanBarron07}. Artsein, Ball, Barthe \& Naor's proof of the EPI~\eqref{abbn3}, which is generalized and simplified by Madiman and Barron, is through an integration of the FII~\eqref{abbnfii} over the path $\{\sqrt{t}\,X+\sqrt{1-t}\,Z\}$ (in\cite{ABBN04}, see~\eqref{i2}) or $\{X+\sqrt{t}\,Z\}$ (in\cite{MadimanBarron07}, see~\eqref{i1}). Tulino and Verd\'u provided the corresponding proof via MMSE\cite{TulinoVerdu06}, through an integration of the MMSE inequality~\eqref{abbnmmsei} over the path $\{\sqrt{t}\,X+Z\}$ (see~\eqref{i3}). Again the approaches corresponding to \eqref{abbnfii} and~\eqref{abbnmmsei} are equivalent by virtue of the complementary relation~\eqref{fimmsewhite}, as explained in section~\ref{sec-verduguo}.

That the MII~\eqref{abbnmii} holds is easily shown through~\eqref{abbnfii} or~\eqref{abbnmmsei} and de Bruijn's identity~\eqref{debruijngauss} or~\eqref{debruijnmmse}.
However, the author was not able to extend the ideas in the proof of Theorem~\ref{th-mii} to provide a direct proof of the MII~\eqref{abbnmii}, which letting $\sigma^2_Z\to\infty$ would yield an easy proof of the generalized EPI~\eqref{abbn3}. 
Such an extension perhaps involves a generalization of the data processing inequality or Sato's inequality in~\eqref{a}, which using the relation $\sum_i a_i X_i=\frac{1}{\sqrt{k}}\sum_S a_S X_S$ would yield the inequality
$I(\sum_i a_i X_i + Z;Z) \leq  \sum_S  I(X_S+a_s Z;Z)+o(\sigma^2_Z)$.

\subsection{EPI, FII and MII for Gas Mixtures}

There is a striking resemblance between the original inequalities~\eqref{epi3},~\eqref{fii3},~\eqref{mmsei} and~\eqref{mii} for linear mixtures of independent random vectors, and known inequalities concerning entropy and Fisher information for linear ``gas mixtures'' of probability distributions.

\begin{proposition}\label{prop-gas}
Let the random variable $I$ have distribution $p(i)=a_i^2$ where $\sum_i a_i^2=1$, let $(X_i)_i$ be finitely many random $n$-vectors independent of $I$, and let $Z$ be white Gaussian, independent of $(X_i)_i$ and $I$. Then
\begin{subequations}\label{gas}
\begin{align}
J(X_I)& \leq \sum_i a_i^2 J(X_i) \label{fiigas}\\
\Var(X_I|X_I+Z)& \geq \sum_i a_i^2 \Var(X_i|X_i+Z) \label{mmsegas}\\
h(X_I) &\geq \sum_i a_i^2 h(X_i) \label{epigas}\\
I(X_I+Z;Z)&\leq \sum_i a_i^2 I(X_i+Z;Z). \label{miigas}
\end{align}
\end{subequations}
\end{proposition}
\medskip

Noting that $X_I$ has distribution $p_{X_I}(x)=\sum_i p(i) p(x|i) = \sum_i a_i^2 p_{X_i}(x)$, the ``FII''~\eqref{fiigas} can be proved directly as follows. Let $S(x)=\nabla p_{X_I}(x)/p_{X_I}(x)$ and $S_i(x)=\nabla p_{X_i}(x)/p_{X_i}(x)$ be the score functions of $X_I$ and the $(X_i)_i$, respectively, and define $\lambda_i(x)=a_i^2 p_{X_i}(x)/p_{X_I}(x)$ for all $i$. Then $\sum_i \lambda_i(x) =1$, $S(x)=\sum_i \lambda_i(x) S_i(x)$ and since the squared norm is convex, $\|S(x)\|^2 \leq \sum_i \lambda_i(x)\|S_i(x)\|^2 = p^{-1}_{X_I}(x) \sum_i a_i^2 p_{X_i}(x) \|S_i(x)\|^2$. Averaging over $p_{X_I}(x)$ gives~\eqref{fiigas}. 

Once~\eqref{fiigas} is established, the conventional techniques presented in Section~\ref{sec-earlier} can be easily adapted to deduce the other inequalities~\eqref{mmsegas}--\eqref{miigas}: Substituting $(X_i+Z_i)_i$ for $(X_i)_i$ in~\eqref{fiigas}, where the $(Z_i)_i$ are independent copies of $Z$, and noting that $X_I+Z_I$ has the same probability distribution as $X_I+Z$, we obtain the inequality $J(X_I+Z) \leq \sum_i a_i^2 J(X_i+Z)$; applying the complementary relation~\eqref{fimmsewhite} gives~\eqref{mmsegas}; integrating using de Bruijn's identity~\eqref{debruijngauss} or~\eqref{debruijnmmse} gives~\eqref{miigas}, from which~\eqref{epigas} follows as in the proof of Theorem~\ref{th-mii}.

In the present case, however,~\eqref{epigas} and~\eqref{miigas} are already well known. In fact, since $p_{X_I}(x) = \sum_i a_i^2 p_{X_i}(x)$ is a convex combination of distributions, the ``EPI''~\eqref{epigas} is nothing but the classical concavity property of entropy, seen as a functional of the probability distribution\cite{Gallager68,McEliece02,CoverThomas06}.
This is easily established by noting that since conditioning decreases entropy, $h(X_I) \geq h(X_I|I)=\sum_i p(i) h(X_i)$. Also the ``MII''~\eqref{miigas} is just the classical convexity of mutual information $I(Y,Z)$, seen as a functional of the distribution $p(y|z)$ for fixed $p(z)$\cite[Thm. 2.7.4]{CoverThomas06}, \cite[Thm. 4.4.3]{Gallager68}, \cite[Thm. 1.7]{McEliece02}.

Accordingly, we may reverse the order of implication and derive the corresponding convexity property of Fisher information~\eqref{fiigas} anew from the ``MII''~\eqref{miigas}. Indeed,~\eqref{miigas} can be rewritten in the form
$$
h(X_I+\sqrt{t}\,Z)-h(X_I)\leq \sum_i a_i^2 \bigl(h(X_i+\sqrt{t}\,Z)-h(X_i)\bigr)\qquad (t>0).
$$ 
Dividing both sides by $t$ and letting $t\to 0$ gives~\eqref{fiigas} by virtue of de Bruijn's identity. This derivation is much shorter than earlier proofs of inequality~\eqref{fiigas}\cite[Lemma 6]{DemboCoverThomas91}, \cite{BudianuTong03}.
The convexity property of Fisher information finds application in channel estimation\cite{BudianuTong05} and thermodynamics\cite{FriedenPlastinoPlastinoSoffer99}.

\section*{Acknowledgments}

The author wishes to thank the Associate Editor for his patience, and the anonymous reviewers for their helpful comments and for pointing out several references and the counterexample of Section~\ref{sec-hdef}. Preliminary advices from Prof. Shlomo Shamai, Prof.~Jun Chen and Dr.~Yihong Wu are also gratefully acknowledged.

\begin{biographynophoto}%[{\includegraphics[width=1in,height=1.25in,clip,keepaspectratio]{./rioul.jpg}}]
{Olivier Rioul}  was born on the fourth of July, 1964.
He received the Dipl. Ing. degree from the Ecole Polytechnique, Palaiseau, France, in 1987, the Dipl. Ing. degree in electrical engineering from the Ecole Nationale Sup\'erieure des T\'el\'ecommunications (ENST), Paris, France, in 1989, and the Ph. D. degree also from the ENST, in 1993.
From 1989 to 1994 he was with the Centre National d'Etudes des T\'el\'ecommunications (CNET) in Issy-les-Moulineaux, France, where he worked on wavelet theory and image compression. In 1994 he joined the ENST, now T\'el\'ecom ParisTech, where he is currently Associate Professor. His research interests include quantization, entropy coding, transform coding, error-control codes, joint source-channel coding and information theory.
\end{biographynophoto} 

\end{document}